\input harvmac
\input epsf
\tolerance=10000



\def\coeff#1#2{\relax{\textstyle {#1 \over #2}}\displaystyle}
\def\clam{\lambda}
\def\ie{{\it i.e.}}

\def\cC{{\cal C}} 
 \def\cG{{\cal G}}
\def\cH{{\cal H}} \def\cI{{\cal I}}
 \def\cK{{\cal K}}
\def\cL{{\cal L}} 
\def\cN{{\cal N}} \def\cO{{\cal O}}
 
 \def\cS{{\cal S}}
 
 \def\cW{{\cal W}}
\def\cX{{\cal X}} 
\def\ev#1{\langle#1\rangle}

\def\IP{\relax{\rm I\kern-.18em P}}
\def\IR{\relax{\rm I\kern-.18em R}}
%

%


\def\sym{  \> {\vcenter  {\vbox
               {\hrule height.6pt
                \hbox {\vrule width.6pt  height5pt
                       \kern5pt
                       \vrule width.6pt  height5pt
                       \kern5pt
                       \vrule width.6pt height5pt}
                \hrule height.6pt}
                          }
               } \>
            }
\def\fund{  \> {\vcenter  {\vbox
               {\hrule height.6pt
                \hbox {\vrule width.6pt  height5pt
                       \kern5pt
                       \vrule width.6pt  height5pt }
                \hrule height.6pt}
                          }
                    } \>
            }
\def\anti{ \>  {\vcenter  {\vbox
               {\hrule height.6pt
                \hbox {\vrule width.6pt  height5pt
                       \kern5pt
                       \vrule width.6pt  height5pt }
                \hrule height.6pt
                \hbox {\vrule width.6pt  height5pt
                       \kern5pt
                       \vrule width.6pt  height5pt }
                \hrule height.6pt}
                          }
               } \>
            }


\lref\dv{
R.~Dijkgraaf and C.~Vafa,
``Matrix models, topological strings, and supersymmetric gauge
theories,''
Nucl.\ Phys.\ B {\bf 644}, 3 (2002), {\tt
arXiv:hep-th/0206255};
``On geometry and matrix models,''
Nucl.\ Phys.\ B {\bf 644}, 21 (2002),
{\tt arXiv:hep-th/0207106};
``A perturbative window into non-perturbative physics,'' {\tt
arXiv:hep-th/0208048};
``N = 1 supersymmetry, deconstruction, and bosonic gauge theories,''
{\tt arXiv:hep-th/0302011}.}
%
\lref\BCOV{
M.~Bershadsky, S.~Cecotti, H.~Ooguri and C.~Vafa,
``Kodaira-Spencer theory of gravity and exact results for quantum
string amplitudes,''
Commun.\ Math.\ Phys.\  {\bf 165}, 311 (1994)
[arXiv:hep-th/9309140].
}
\lref\Oovc{
H.~Ooguri and C.~Vafa,
``The C-deformation of gluino and non-planar diagrams,''
arXiv:hep-th/0302109.
H.~Ooguri and C.~Vafa,
``Gravity induced C-deformation,''
arXiv:hep-th/0303063.
}
%
\lref\CachazoRY{
F.~Cachazo, M.~R.~Douglas, N.~Seiberg and E.~Witten,
``Chiral rings and anomalies in supersymmetric gauge theory,''
JHEP {\bf 0212}, 071 (2002),
{\tt arXisv:hep-th/0211170}.
}
%
\lref\KrausJF{ P.~Kraus and M.~Shigemori,
``On the matter of the
Dijkgraaf-Vafa conjecture,'' arXiv:hep-th/0303104.
}
\lref\AldayGB{
L.~F.~Alday and M.~Cirafici,
``Effective superpotentials via Konishi anomaly,''
arXiv:hep-th/0304119.
}
\lref\KrausJV{
P.~Kraus, A.~V.~Ryzhov and M.~Shigemori,
``Loop equations, matrix models, and N = 1 supersymmetric gauge
theories,''
arXiv:hep-th/0304138.
}

%
\lref\VafaWI{ C.~Vafa,
``Superstrings and topological strings at
large N,'' J.\ Math.\ Phys.\  {\bf 42}, 2798 (2001)
[arXiv:hep-th/0008142].
}
\lref\CachazoJY{ F.~Cachazo, K.~A.~Intriligator and C.~Vafa, ``A
large N duality via a geometric transition,'' Nucl.\ Phys.\ B {\bf
603}, 3 (2001) [arXiv:hep-th/0103067].
}
\lref\DijkgraafXD{ R.~Dijkgraaf, M.~T.~Grisaru, C.~S.~Lam,
C.~Vafa and D.~Zanon,
``Perturbative computation of glueball
superpotentials,'' arXiv:hep-th/0211017.
}
\lref\witr{ E.~Witten,
``Chiral ring of Sp(N) and SO(N)
supersymmetric gauge theory in four  dimensions,''
arXiv:hep-th/0302194.
}
\lref\bjsv{
M.~Bershadsky, A.~Johansen, V.~Sadov and C.~Vafa,
``Topological reduction of 4-d SYM to 2-d sigma models,''
Nucl.\ Phys.\ B {\bf 448}, 166 (1995)
[arXiv:hep-th/9501096].
}
\lref\smh{
J.~A.~Harvey, G.~W.~Moore and A.~Strominger,
``Reducing S duality to T duality,''
Phys.\ Rev.\ D {\bf 52}, 7161 (1995)
[arXiv:hep-th/9501022].
}
%
\lref\HarveyTG{
J.~A.~Harvey, G.~W.~Moore and A.~Strominger,
``Reducing S duality to T duality,''
Phys.\ Rev.\ D {\bf 52}, 7161 (1995)
[arXiv:hep-th/9501022].
}
\lref\hv{ K.~Hori and C.~Vafa, ``Mirror symmetry,''
arXiv:hep-th/0002222.
}
\lref\loijenga{E. Looijenga,``Root Systems And Elliptic Curves,''
Invent.
Math. {\bf 38}, 17 (1977)\semi ``Invariant Thoery For Generalized Root
Systems,'' Invent. Math. {\bf 61}, 1 (1980).}
\lref\wiind{E.~Witten, ``Constraints On Supersymmetry Breaking,''
Nucl.\ Phys.\ B {\bf 202}, 253 (1982).
``Toroidal compactification without vector structure,'' JHEP {\bf
9802}, 006 (1998) [arXiv:hep-th/9712028].
}
\lref\kv{ S.~Katz and C.~Vafa, ``Geometric engineering of N = 1
quantum field theories,'' Nucl.\ Phys.\ B {\bf 497}, 196 (1997)
[arXiv:hep-th/9611090].
}
%
\lref\KawaiYF{
H.~Kawai, T.~Kuroki and T.~Morita,
``Dijkgraaf-Vafa theory as large-N reduction,''
arXiv:hep-th/0303210.
}
\lref\dohol{ N.~Dorey, T.~J.~Hollowood and S.~Prem Kumar, ``An
exact elliptic superpotential for N = 1* deformations of finite  N
= 2 gauge theories,'' Nucl.\ Phys.\ B {\bf 624}, 95 (2002)
[arXiv:hep-th/0108221].
}
\lref\Doreyb{ N.~Dorey, T.~J.~Hollowood, S.~P.~Kumar and
A.~Sinkovics, ``Massive vacua of N = 1* and S-duality from matrix
models,'' JHEP {\bf 0211}, 040 (2002) [arXiv:hep-th/0209099].
}
\lref\Doreya{ N.~Dorey, T.~J.~Hollowood, S.~Prem Kumar and
A.~Sinkovics, ``Exact superpotentials from matrix models,'' JHEP
{\bf 0211}, 039 (2002) [arXiv:hep-th/0209089].
}
%
\lref\wittog{
E.~Witten,
``On The Structure Of The Topological Phase Of Two-Dimensional
Gravity,''
Nucl.\ Phys.\ B {\bf 340}, 281 (1990).
}
\lref\AffleckMK{
I.~Affleck, M.~Dine and N.~Seiberg,
``Dynamical Supersymmetry Breaking In Supersymmetric QCD,''
Nucl.\ Phys.\ B {\bf 241}, 493 (1984).
}
\lref\IntriligatorJR{
K.~A.~Intriligator, R.~G.~Leigh and N.~Seiberg,
``Exact superpotentials in four-dimensions,''
Phys.\ Rev.\ D {\bf 50}, 1092 (1994)
[arXiv:hep-th/9403198].
}
\lref\av{
M.~Aganagic and C.~Vafa,
``Perturbative derivation of mirror symmetry,''
arXiv:hep-th/0209138.
}
\lref\IntriligatorID{
K.~A.~Intriligator and N.~Seiberg,
``Duality, monopoles, dyons, confinement and oblique confinement in
supersymmetric SO(N(c)) gauge theories,''
Nucl.\ Phys.\ B {\bf 444}, 125 (1995)
[arXiv:hep-th/9503179].
}
\lref\IntriligatorNE{
K.~A.~Intriligator and P.~Pouliot,
``Exact superpotentials, quantum vacua and
duality in supersymmetric SP(N(c)) gauge theories,''
Phys.\ Lett.\ B {\bf 353}, 471 (1995)
[arXiv:hep-th/9505006].
}

\lref\LeighQP{
R.~G.~Leigh and M.~J.~Strassler,
``Duality of Sp(2N(c)) and S0(N(c)) supersymmetric gauge theories
 with adjoint matter,''
Phys.\ Lett.\ B {\bf 356}, 492 (1995)
[arXiv:hep-th/9505088].
}

\lref\IntriligatorAX{
K.~A.~Intriligator, R.~G.~Leigh and M.~J.~Strassler,
``New examples of duality in chiral and nonchiral supersymmetric gauge
 theories,''
Nucl.\ Phys.\ B {\bf 456}, 567 (1995)
[arXiv:hep-th/9506148].
}

\lref\IntriligatorFF{
K.~A.~Intriligator,
``New RG fixed points and duality in supersymmetric SP(N(c)) and SO(N(c))
 gauge theories,''
Nucl.\ Phys.\ B {\bf 448}, 187 (1995)
[arXiv:hep-th/9505051].
}

\lref\CsakiVV{
C.~Csaki and H.~Murayama,
``Instantons in partially broken gauge groups,''
Nucl.\ Phys.\ B {\bf 532}, 498 (1998)
[arXiv:hep-th/9804061].
}
\lref\CachazoPR{
F.~Cachazo and C.~Vafa,
``N = 1 and N = 2 geometry from fluxes,''
arXiv:hep-th/0206017.
}
\lref\Intstr{
E.~J.~Martinec and N.~P.~Warner,
``Integrable systems and supersymmetric gauge theory,''
Nucl.\ Phys.\ B {\bf 459}, 97 (1996)
[arXiv:hep-th/9509161].
\hfill \break
R.~Donagi and E.~Witten,
``Supersymmetric Yang-Mills Theory And Integrable Systems,''
Nucl.\ Phys.\ B {\bf 460}, 299 (1996)
[arXiv:hep-th/9510101].
\hfill \break
E.~J.~Martinec,
``Integrable Structures in Supersymmetric Gauge and String Theory,''
Phys.\ Lett.\ B {\bf 367}, 91 (1996)
[arXiv:hep-th/9510204].
}
%
\lref\SeibergRS{
N.~Seiberg and E.~Witten,
``Electric - magnetic duality, monopole condensation, and confinement
in N=2 supersymmetric Yang-Mills theory,''
Nucl.\ Phys.\ B {\bf 426}, 19 (1994)
[Erratum-ibid.\ B {\bf 430}, 485 (1994)]
[arXiv:hep-th/9407087].
}
 \lref\Seiberg{
N.~Seiberg,
``Exact results on the space of
vacua of four-dimensional SUSY gauge theories,''
Phys.\ Rev.\ D {\bf 49}, 6857 (1994)
[arXiv:hep-th/9402044].
}
\lref\losev{ A.~Losev,
``Descendants Constructed From Matter Field
In Topological Landau-Ginzburg Theories Coupled To Topological
Gravity,'' Theor.\ Math.\ Phys.\  {\bf 95}, 595 (1993) [Teor.\
Mat.\ Fiz.\  {\bf 95}, 307 (1993)] [arXiv:hep-th/9211090].
}
\lref\dgkv{ R.~Dijkgraaf, S.~Gukov, V.~A.~Kazakov and C.~Vafa,
``Perturbative analysis of gauged matrix models,''
arXiv:hep-th/0210238.
}
\lref\KlemmCY{
A.~Klemm, K.~Landsteiner, C.~I.~Lazaroiu and I.~Runkel,
``Constructing gauge theory geometries from matrix models,''
arXiv:hep-th/0303032.
}
\lref\AshokBI{
S.~K.~Ashok, R.~Corrado, N.~Halmagyi, K.~D.~Kennaway and
C.~Romelsberger,
``Unoriented strings, loop equations, and N = 1 superpotentials from
matrix models,''
arXiv:hep-th/0211291.
}
\lref\oz{
H.~Ita, H.~Nieder and Y.~Oz,
``Perturbative computation of glueball superpotentials for SO(N) and
USp(N),''
JHEP {\bf 0301}, 018 (2003)
[arXiv:hep-th/0211261].
}
\lref\vato{
C.~Vafa,
``Brane/anti-brane systems and U(N$|$M) supergroup,''
arXiv:hep-th/0101218.
}
\lref\gid{
S.~B.~Giddings and J.~M.~Pierre,
``Some exact results in supersymmetric theories based on exceptional
groups,''
Phys.\ Rev.\ D {\bf 52}, 6065 (1995)
[arXiv:hep-th/9506196].
}
\lref\rosz{
A.~Brandhuber, H.~Ita, H.~Nieder, Y.~Oz and C.~Romelsberger,
``Chiral rings, superpotentials and the vacuum structure of N = 1
supersymmetric gauge theories,''
arXiv:hep-th/0303001.
}
%
\lref\BoelsFH{
R.~Boels, J.~de Boer, R.~Duivenvoorden and J.~Wijnhout,
``Nonperturbative superpotentials and compactification to three
dimensions,''
arXiv:hep-th/0304061.
}
\lref\klebw{
I.~R.~Klebanov and E.~Witten,
``Superconformal field theory on threebranes at a Calabi-Yau
singularity,''
Nucl.\ Phys.\ B {\bf 536}, 199 (1998)
[arXiv:hep-th/9807080].
I.~R.~Klebanov and E.~Witten,
``AdS/CFT correspondence and symmetry breaking,''
Nucl.\ Phys.\ B {\bf 556}, 89 (1999)
[arXiv:hep-th/9905104].
}
\lref\klst{
I.~R.~Klebanov and M.~J.~Strassler,
``Supergravity and a confining gauge theory: Duality cascades and
chiSB-resolution of naked singularities,''
JHEP {\bf 0008}, 052 (2000)
[arXiv:hep-th/0007191].
}
\lref\LosevTU{
A.~Losev, N.~Nekrasov and S.~L.~Shatashvili,
``Freckled instantons in two and four dimensions,''
Class.\ Quant.\ Grav.\  {\bf 17}, 1181 (2000)
[arXiv:hep-th/9911099].
}
\lref\aha{O. Aharony, as quoted in C.~Vafa,
``Puzzles at large N,''
arXiv:hep-th/9804172.  }
\lref\SW{
N.~Seiberg and E.~Witten,
``Gauge dynamics and compactification to three dimensions,''
arXiv:hep-th/9607163.
}
%
\lref\DoreySJ{
N.~Dorey,
``An elliptic superpotential for softly broken N = 4 supersymmetric
Yang-Mills theory,''
JHEP {\bf 9907}, 021 (1999)
[arXiv:hep-th/9906011].
}
\lref\veya{
G.~Veneziano and S.~Yankielowicz,
``An Effective Lagrangian For The Pure N=1 Supersymmetric Yang-Mills
Theory,''
Phys.\ Lett.\ B {\bf 113}, 231 (1982).
}
\lref\TaylorBP{
T.~R.~Taylor, G.~Veneziano and S.~Yankielowicz,
``Supersymmetric QCD And Its Massless Limit: An Effective Lagrangian Analysis,''
Nucl.\ Phys.\ B {\bf 218}, 493 (1983).
}
 \lref\bmrt{I. Bena, H. Murayama, R. Roiban
and R. Tatar,``Matrix Model Description of Baryonic Deformations,''
[arXiv:hep-th/0303115].
}
%
\lref\ElitzurHC{
S.~Elitzur, A.~Giveon, D.~Kutasov, E.~Rabinovici and A.~Schwimmer,
``Brane dynamics and N = 1 supersymmetric gauge theory,''
Nucl.\ Phys.\ B {\bf 505}, 202 (1997)
[arXiv:hep-th/9704104].
}
\lref\CachazoSG{
F.~Cachazo, B.~Fiol, K.~A.~Intriligator, S.~Katz and C.~Vafa,
``A geometric unification of dualities,''
Nucl.\ Phys.\ B {\bf 628}, 3 (2002)
[arXiv:hep-th/0110028].
}
%
\lref\NaculichPQ{
S.~G.~Naculich, H.~J.~Schnitzer and N.~Wyllard,
``A cascading N = 1 Sp(2N+2M) x Sp(2N) gauge theory,''
Nucl.\ Phys.\ B {\bf 638}, 41 (2002)
[arXiv:hep-th/0204023].
}
%
\lref\ChoBI{
P.~L.~Cho and P.~Kraus,
``Symplectic SUSY gauge theories with antisymmetric matter,''
Phys.\ Rev.\ D {\bf 54}, 7640 (1996)
[arXiv:hep-th/9607200].
}
\lref\CsakiEU{
C.~Csaki, W.~Skiba and M.~Schmaltz,
``Exact results and duality for
Sp(2N) SUSY gauge theories with an  antisymmetric tensor,''
Nucl.\ Phys.\ B {\bf 487}, 128 (1997)
[arXiv:hep-th/9607210].
}
%
\lref\WittenHF{
E.~Witten,
Commun.\ Math.\ Phys.\  {\bf 121}, 351 (1989).
}
%

\Title{\vbox{
\hbox{CALT-68-2437}
\hbox{HUTP-03/A030}
\hbox{USC-03/03}
\hbox{UCSD-PTH-03-06}
\hbox{\tt hep-th/0304271}  \vskip -0.5cm}}
{\vbox{\vskip -0.5cm
\centerline{\hbox{The Glueball Superpotential}} }}

\vskip 0.1in
\centerline{Mina Aganagic$^1$, Ken Intriligator$^2$, Cumrun
Vafa$^{1,3}$  and
Nicholas P. Warner $^4$}
\vskip .2in

\centerline{$^1$ Jefferson Physical Laboratory, Harvard University,
Cambridge, MA 02138, USA}
\vskip .05in
\centerline{$^2$ Department of Physics, University of California, San
Diego,
La Jolla, CA 92093-0354, USA}
\vskip .05in
\centerline{$^3$ California Institute of Technology 452-48,
Pasadena, CA 91125, USA}
\vskip .05in
\centerline{$^4$  Department of Physics and Astronomy, USC, Los Angeles,
CA 90089-0484, USA}

\vskip .1in
\centerline{{\bf Abstract}}
\medskip

We compute glueball superpotentials for four-dimensional, ${\cal N}=1$
supersymmetric gauge theories, with arbitrary gauge groups and massive
matter representations.  This is done by perturbatively integrating
out massive charged fields.  The Feynman diagram computations
simplify, and are related to the corresponding matrix model.  This
leads to a natural notion of ``projection to planar diagrams'' for
arbitrary gauge groups and representations.  We discuss a general
ambiguity in the glueball superpotential $W(S)$ for terms, $S^n$,
whose order, $n$ is greater than the dual Coxeter number.  This
ambiguity can be resolved for all classical gauge groups $(A,B,C,D)$,
via a natural embedding in an infinite rank supergroup. We use this to
resolve some recently raised puzzles. For exceptional groups, we
compute the superpotential terms for low powers of the glueball field
and propose an all-order completion for some examples including ${\cal
N}=1^*$ for all simply-laced groups. We also comment on
compactification of these theories to lower dimensions.

\vskip .2in


 \Date{\sl {April, 2003}}

\newsec{Introduction}

New insights have recently been obtained into the non-perturbative
dynamics of supersymmetric gauge theories in four and higher
dimensions.  It has been found that perturbative computation of the glueball
superpotential in four-dimensional supersymmetric gauge theories admitting a
large $N$ description is related to matrix model amplitudes.  Moreover, the
extremization of the superpotential provides exact, non-perturbative
 results for the gauge
theory \dv .    This connection between the superpotential generated
by integrating out massive fields and matrix model amplitudes
has been explicitly demonstrated in \DijkgraafXD .  In particular, 
using supergraph techniques, which mirror the corresponding topological
superstring computations  \refs{\BCOV, \Oovc}, one can see
a dramatic simplification of the Feynman amplitudes contributing to the
 glueball
superpotential.

In this paper we follow the strategy of \DijkgraafXD\ to compute the
exact  glueball
superpotential for general gauge group and massive matter content.  We
find that the dramatic simplifications occur for {\it any} group and
{\it any} massive representation.   Moreover this leads to an interesting
interpretation of the ``planar projection'' for arbitrary groups and
 representations.

To simplify the analysis, as in \DijkgraafXD, we
use constant abelian gluino backgrounds.
For such backgrounds one can compute the glueball superpotential
$W(S)$ for terms $S^n$ with $n\leq $rank$(G)$ \foot{One
can in principle develop this further to compute terms $S^k$ with $r<k<h$ by
including some non-abelian configurations.}.
A general aspect of the glueball superpotential $W(S)$, which we
discuss in detail, is that it is ambiguous at terms of order $S^n$
with $n\geq h$, the dual Coxeter number.  These higher order terms
depend on the details of how the theory is defined in the
ultraviolet, the ``UV completion of the theory."  These ambiguities
have to do with instantons:  Specifically,  $\ev{S^h}$ is classically zero, but
is non-zero in the quantum theory because of instantons.  On the other hand,
terms of order $S^n$, with $n<h$, generally do not depend on the UV details
 of the
theory; these unambiguous terms correspond to fractional instanton effects.

We propose a natural completion for these F-terms for general
classical gauge groups following the last reference in \dv\ which
leads to an all order superpotential for $S$.  This prescription is
first motivated via a general expression derived here for the
contribution of any Feynman graph to the superpotential.  This
expression is valid for any gauge theory with arbitrary matter
content, and it leads to to an F-term completion via analytic
continuation in $N$ for the classical groups.  However, there is a
more physical, refined and indeed predictive realization of this
F-term completion in terms of supergroups. 

The group $G(N)= U(N)$, or $SO(N)$, or $Sp(N)$ is completed into the
supergroup $G(N+k|k)$, with large $k$.  This is very natural from the
viewpoint of branes/anti-brane systems \vato: We add a large number
$k$ of brane anti-brane pairs. For example, we embed $U(N)$ into
$U(N+k|k)$, and in the latter theory the glueball superpotential is
unambiguous up to order $S^{N+k}$.  The supertrace structure ensures
that the coefficients in the $U(N+k|k)$ glueball superpotential are
independent of $k$, and so it is natural to compute the $U(N)$
glueball superpotential, including terms $S^n$ with $n$ arbitrarily
large, by going to the $U(N+k|k)$ theory with $k\rightarrow \infty$.
The $G(N+k|k)$ supergroup thus leads to a natural ``F-term
completion'' of $G(N)$, in the sense that the $G(N)$ theory can be
viewed as a Higgs branch of $G(N+k|k)$ theory, which is F-term
complete as $k\rightarrow \infty$.  For large $k$, this F-completion
of the glueball superpotential is computable, as in \DijkgraafXD ,
perturbatively to arbitrarily high order, it is exact and this higher
power completion is consistent with a generalized Konishi anomaly
\CachazoRY.

In some instances the $G(N+k|k)$ F-completion differs from some of the standard
UV completions of $G(N)$.  As we discuss, this generally happens when
there are residual instanton effects associated with the Higgsing of
$G(N+k|k)$ to $G(N)$.  These effects arise  when the
quotient  $G(N+k|k)/G(N)$ has the appropriate topology to give rise
to instantons.  We will analyze precisely when our $G(N+k|k)$ F-completions
differ from the standard $G(N)$ UV completions, and how this explains apparent
discrepancies between the matrix model results and standard gauge theory
results.
For example, standard $U(1)$ gauge theory does not have a glueball
superpotential,
but its F-completion into $U(1+k|k)$ does, with the difference coming
{}from instantons in $U(2|1)/U(1)$.  This issue of the F-completion
could explain the apparent discrepancy observed in \KrausJF\ between
the matrix model/perturbative glueball superpotential results
(equivalently, the generalized Konishi anomaly results \refs{\AldayGB,
\KrausJV}) and standard supersymmetric gauge theory results.

Gauge theories based upon supergroups are non-unitary, and
so this might seem to be a rather unphysical way to resolve   F-term
ambiguities.
However one should note that F-terms of broken gauge systems
will inovolve ghost-like chiral fields as shown in \dgkv.  Moreover,
gauge theories for
non-unitary supergroups can have the same F-terms as unitary quiver theories,
as demonstrated in the last reference in \dv.  Thus, one cannot distinguish
unitary and non-unitary gauge theories based solely upon F-terms.
It is therefore natural to
extend the class of theories whose F-terms are to be studied so as
to include both
unitary and non-unitary gauge systems.

It is also important to find F-completions of the glueball
superpotential for non-classical
groups.  We point out at least two ways this may be done:
If the theory has a branch
with classical groups emerging as unbroken groups then the foregoing
prescription leads to
an answer.   In some other instances, like ${\cal N}=1^*$ theories,
the amplitudes depend upon
group theoretic factors in a universal way and one may ``analytically
continue'' in the choice of
the group to define an  F-completion.   In particular, we use our general expression
for the diagrammatic contributions to obtain the superpotential to three loops
for the ${\cal N}=1^*$ theories for any simply-laced gauge group.
This leads us to conjecture
an analytic continuation prescription in which $N S^\ell$ is replaced by
$S^\ell \sum (p_a)^\ell$,
where $p_a$ are the extended Dynkin labels of the extending root of the
underlying algebra.   This conjecture is further supported by considering
compactifications to two and three dimensions.

We discuss the relationship between $\cN=1$ supersymmetric gauge
theories in four dimensions
and a particular class of
 $\cN=2$ supersymmetric sigma models in two dimensions.
The latter is obtained via the moduli space of flat connections on
the compactifying torus.
One can then see a direct relationship between the instanton corrections
to the chiral rings of both
theories.  This approach provides some insights as to how the ambiguities could
be related to ``gravitational descendants'' in the two-dimensional theory.
On a more
straightforward level, in $\cN=1^*$ theories the mirror of this sigma model
is naturally related to
the integrable structures used in  \refs{\Intstr, \DoreySJ, \BoelsFH}.  In particular,
this enables us to
recompute the diagrammatic expansion of the glueball superpotential by making
 a duality transformation of the elliptic Calogero-Moser
superpotentials of \DoreySJ. We show that this agrees
with the direct diagrammatic computation of the glueball superpotential.

The organization of this paper is as follows:  In section 2 we
discuss some general aspects of the glueball superpotential.  In
section 3 we compute the corrections to the superpotential for arbitrary
groups and representation up to the
glueball field to the power of the rank of the gauge group.  One
can make sense of the notion
of the ``projection to planar diagrams''  for arbitrary groups
provided the number of loops is less than the rank.  In section 4 we propose
the natural F-completion of our theories.
We argue that the matrix model results
should be understood as referring to this particular F-completion.
In section 5 we discuss some special cases, where our F-completion
differs from more standard UV definitions of some gauge theories,
at the non-perturbative level.  The difference comes from instantons
in the partially broken group $G(N+k|k)/G(N)$.  In section 6 we
discuss some further glueball superpotential examples
and compute the glueball superpotential for the ${\cal N}=1^*$ theory
for arbitrary simply-laced gauge groups, to arbitrarily high order  in terms
of a proposed F-completion.
In section 7 we consider compactifications to 2 and 3 dimensions
and the meaning of the superpotential computation in these cases.
In section 8 we make some final remarks.
In appendix A we derive a group theory result that
we need for the Feynman diagram computations.

{\bf Note added in revised version, Nov. 2003:}

In the original version of this paper, we speculated that the residual
instantons effects discussed in this paper, associated with the F-completion, should 
resolve the apparent discrepancy of \KrausJF\ between the matrix model
and standard gauge theory.    We now know that this speculation was incorrect.
We still claim that the matrix model refers to the $G(N+k|k)$ F-completion,
and that the results thus obtained could, in principle, differ from standard
gauge theory by residual instanton effects.  But the matrix model side of the
computation must be done appropriately, which requires glueball fields for
$U(1)$ and $Sp(0)$ factors.  Following the first version of the present paper,
\ref\CachazoKX{
F.~Cachazo,
``Notes on supersymmetric Sp(N) theories with an antisymmetric tensor,''
arXiv:hep-th/0307063.
}\
appeared, which gave a particular treatment of the $Sp(0)$ factors 
introduced here; this treatment was later explained from the string theory
perspective, and extended to a general prescription for all low-rank classical
groups in 
\ref\IKRSV{
K.~Intriligator, P.~Kraus, A.~V.~Ryzhov, M.~Shigemori and C.~Vafa,
``On Low Rank Classical Groups in String Theory, Gauge Theory and Matrix Models,''
arXiv:hep-th/0311181.
}.
Upon redoing the matrix model computation according to this new prescription,
the results agree perfectly with standard gauge theory \refs{\CachazoKX, \IKRSV}.
This is consistent with the claim made here that the matrix model refers to the $G(N+k|k)$
completion, because further investigation of the theories of sect. 5.4 reveals that the
residual instanton type effects, which in principle could have spoiled the agreement
with standard gauge theory, can -- and here do --  exhibit remarkable cancellations
\IKRSV.

\newsec{The Glueball Superpotential}

The glueball superfield of a four-dimensional, ${\cal N}=1$ supersymmetric
gauge theory is defined by
\eqn\Sdefn{S=\epsilon^{\alpha \beta} \,  g_{AB} \, {\cal W}^A_\alpha \,
   {\cal W}^B_{\beta}
\,. }
where $A$ labels the Lie algebra elements, $g_{AB}$ is the
corresponding group invariant inner product and ${\cal W}^A_\alpha$ is
the gluino field.  The central idea in the proposal of \dv\ in gaining
a perturbative window into non-perturbative dynamics of ${\cal N}=1$
supersymmetric gauge theories has been to compute the
glueball superpotential $W(S)$, perturbatively, by integrating out
massive fields.  One then treats $S$ as a good order parameter in the
IR physics and extremizes $W(S)$:
$${dW\over dS}=0\, .$$
This yields, through the values of the superpotential $W$
at the extrema, non-perturbative information about the gauge theory.
This can also be extended to an exact computation of the coupling
constants $\tau _{ij}$ for abelian factors.

For a pure gauge theory with dual Coxeter number $h$, the leading
piece of the superpotential is given by the Veneziano-Yankielowicz
superpotential  \veya:
\eqn\vyw{W_{VY}(S)=h\,  S\,( ({\rm log}( S/\Lambda ^3) ~-~  1) ~+~ \tau
S\,,}
where $\tau$ is the gauge coupling constant at scale $\Lambda$.
Upon extremization this yields
\eqn\swis{S=\Lambda^3 \, {\rm e}^{-\tau/h} \,,  \qquad W=-h
\,\Lambda^3\,
{\rm e}^{-\tau/h}\,.}
Note that this is consistent with the breaking
of the non-anomalous ${\bf Z}_{2h}$  subgroup of  the $U(1)_R$ symmetry.
That is, the ${\bf Z}_{2h}$ symmetry acts on $S$ via:
$$S\rightarrow S \, {\rm e}^{-2\pi i/2h}\,, $$
and is broken to ${\bf Z}_2$ by the vacuum expectation value
of the glueball field.

In general, upon integrating out massive charged fields one finds
corrections to the glueball superpotential.  In particular, as
argued in \DijkgraafXD , an $\ell$-loop diagram
involving charged matter fields can contribute a term
$$\delta W(S)~=~ c_\ell \, S^\ell \,.$$

For high powers of $S$ there is an
ambiguity in the definition of $W(S)$.  Classically, $S$ is a bilinear
fermionic fields, and so if we raise $S$ to a large enough power it
will vanish.  In particular $S^k=0$ for $k> {\rm dim}(G)$.  In a quantum
theory it is natural to define powers of $S$ by point splitting or
smearing, and so we could instead consider a smeared glueball field
$$S_\rho (x)=\int d^4x' \rho_x (x') S(x') \,, $$
where $\rho_x(x')$ is a positive smearing function centered
at $x$ with $\int
\rho_x(x') d^4 x'=1$.  In the limit $\rho$ becomes a $\delta$-function
the smeared glueball field goes back to being the ordinary glueball
field $S_\rho(x) \rightarrow S(x)$.  There is {\it a
priori} no reason for $S_{\rho}^k(x)$ to vanish for any $k$.  Now
consider an ${\cal N}=1$ theory and change the UV action by turning
on, by hand, a superpotential term
\eqn\dact{\delta (action)~=~ \int d^4 x\,  d^2\theta \  \sum_{k\geq d}
   a_k\,  S_\rho^k .}
If $d={\rm dim} (G)+1$ then this term disappears in the limit
$\rho_x(x') \rightarrow
\delta (x-x')$.  In fact one can say something stronger:
As was discussed in \CachazoRY\ classically one expects
$$S^k~=~ 0$$
for $k\geq h$.  This equality is a statement in the chiral ring, that is,
one actually has the {\it classical} relation:
\eqn\vanrel{S^k~=~{\overline D}{\cal O}_k \quad {\rm for}\quad   k\geq h
\,,}
for some operators ${\cal O}_k$.
Moreover, the difference $S(x)-S(y)$ is also trivial in the chiral
ring.  Equation \vanrel\ was established for $U(N)$ in \CachazoRY\ and
for $SO(N)$ and $Sp(N)$ in \witr.  In section 7 we will give a further
argument in support of this classical ring relation for any group.

Note that if we have a chirally trivial operator then adding
it to the superpotential does not change the action.
This implies that if we include
deformations with monomials $S^k$ with $k\geq h$ (\ie\ set $d=h$ in
\dact ) in the superpotential, even with the smearing turned on,
the action does not change classically.  However, the quantum theory
{\it will change} through such deformations because the classical
chiral ring relation receives quantum corrections and $S^k$ is no
longer zero for $k\geq h$. This is apparent from the VY
superpotential, which leads to $S\not=0$, and so all the additional
higher power terms in $S$ will be relevant for the IR physics.  Two
${\cal N}=1$ theories can agree classically, but differ quantum
mechanically by superpotential terms involving $S^k$ with $k\geq h$.
Note that for arbitrary addition of $S^k$ with $k\geq h$ there may not
be any corresponding UV complete theory. On the other hand, there
could be several UV complete theories which agree classically, but
differ quantum mechanically by such additions to the
superpotential. Thus, in principle, specifying a classical description
of an ${\cal N}=1$ theory is not enough to determine all the F-terms
unambiguously. This means that there is an inherent ambiguity in what
one means by the quantum ${\cal N}=1$ supersymmetric gauge theory.

The ambiguities set in at instanton number one, because $S^h\sim
e^{-\tau}$.  This means that if two quantum theories are classically
the same then they will have same superpotential, $W$, for the
fractional instantons.  In particular, computations of the value of
the superpotential, $W(q)$, where $q=e^{-\tau/h}$, lead to an
unambiguous answer for all $q^k$ for $k<h$, but ambiguities can begin
to show up at order $q^h$.  That is, if $W_1$ and $W_2$ are the
superpotentials of two theories with the same classical form, then:
$$W_1-W_2=\sum_{s\geq 0} c_s \, q^{s+h}$$
for some constants, $c_s$.   The existence of such
ambiguities  has been noted recently in checking the matrix
model proposal for computation of ${\cal N}=1$ F-terms in
\refs{\Doreya, \KrausJF}.
The obvious question is how to remove the quantum ambiguities?

Sometimes extra symmetry restricts the ambiguities:
For example,  for  pure Yang-Mills theory, insisting on
a non-anomalous
${\bf Z}_{2h}$ symmetry allows only the addition of the terms
$$\Delta W= \sum_n a_n\,  S^{1+nh}$$
to the superpotential. Note that if we add $\Delta W$ to the VY
superpotential the vacuum structure does not qualitatively change.  Or
in fact, as was argued in \refs{\veya , \TaylorBP} , if we impose the structure of the
anomalous $U(1)$ R-symmetry in the superpotential (\ie\ that the
chiral $U(1)$ rotation by $\alpha$ leads to $\delta W =i~\alpha h
S$)\foot{Note that this result only uses the Adler-Bell-Jakiw anomaly
and {\it does not} assume confinement.  However to obtain chiral
symmetry breaking, one has to make the non-trivial assumption that the
glueball field $S$ is a good order parameter for the IR physics.  Note
that there are theories for which there is no confinement but $S$ is
still a good order parameter, such as $U(N)$ with an adjoint field
broken to $U(1)^N$.}  we can rule out any additional higher powers to
the Veneziano-Yankielowicz potential.
This is, however, a rare situation with a high degree of
symmetry. For more general theories with less symmetries this is not
possible.

One approach to fix the ambiguities would be to start with
a conformal fixed point in the UV and flow down to the IR by the
addition of some relevant operators.
However, even in such cases there is room for ambiguity to develop in
definitions of the relevant operators due to operator mixing.  Such
a possibility was already pointed out in \Doreya\ in the context of
mass deformations of ${\cal N}=4$ theories.

Thus one should not look for a unique IR answer, as it would depend on
how the UV completion is achieved.  For theories such as ${\cal N}=1$,
$U(N)$ supersymmetric gauge theories with one adjoint matter multiplet
and arbitrary superpotential, the string theory embedding naturally
provides a UV completion giving unambiguous higher order terms
\refs{\VafaWI, \CachazoJY} thus leading to the matrix model proposal
in \dv .  This was further extended to give an unambiguous proposal
for arbitrary classical groups admitting large $N$ description in \dv
. The idea, motivated from string theory, basically reduces to
computing the glueball superpotential $W$ for the classical group
admitting large $N$ description by taking the large $N$ limit to be
exact for the computation of the glueball superpotential\foot{ This
structure was anticipated from string theory where on the large $N$
gravitational dual, the glueball superpotential is exact at genus
zero.}.  In other words, to compute $S^k$ for any fixed $k$, consider a
sequence of theories indexed by $N$ and take $N$ large enough and
compute analytic expressions for $a_k(N) S^k$ and then substitute a
finite value of $N$ for $a_k(N)$ at the end of the computation.  This
gives an unambiguous completion of all F-terms, and this, in turn
gets related to planar diagrams of the associated matrix model.  Note
that any other completion would lead to differences of order
$q^h=O(e^{-N})$.  In other words the computation of $W$ at the
extremum for any two possible UV completions would lead to the same
exact result to all orders in the $(1/N)$ expansion.

The fact that classically $S^{h+n}=0$ for any non-negative $n$,
implies that perturbatively it makes sense only to compute the
glueball superpotentials for powers up to $h$, that is, $S^k$ with
$k<h$.  Beyond this, the perturbative computation is ambiguous.  In
other words we can compute, in principle, only a truncation of $W$
unambiguously. We will call this the {\it reduced} superpotential,
$W_R$.  The higher powers of the glueball field lead to what one means
quantum mechanically by the corresponding theory. We denote the
corresponding piece of the superpotential consisting of terms with
$S^{n+h}$ with $n\geq 0$, by $W_A$. Thus any $W_A$ can in principle
arise in a quantum theory and should be viewed as part of the quantum
definition of the theory. Put another way, we have the decomposition
$$W(S)=W_R(S)+W_A(S)$$
where $W_R$ is unambiguously computable by integrating out matter fields
and $W_A$ is part of the definition of the quantum theory.
Thus a conservative generalization of the proposal of \dv\ reduces
to the statement that non-perturbative fractional
instanton effects can be computed unambiguously from a perturbative
definition of the theory.  The rest
can also be computed if we know the precise choice of the
non-perturbative
F-completion of the theory.  For classical groups, even
those which do not have a large $N$, `t Hooft description,
we can follow the approach of \dv\ in defining a quantum
completion by considering the large $N$ regularization
of the superpotential computation and substituting finite
$N$ in the analytic computations at the end.  We will discuss the
meaning of such a prescription in section 4.

One aim of this paper is to compute $W_R$ using perturbative
techniques. We will use the technique of \DijkgraafXD\ which considers
a constant, abelian gluino backgrounds.  For such backgrounds, $S^k$
is non-zero for $k\leq r$ where $r$ is the rank of the group, so one
can compute all monomials in $W_R(S)$ up to $S^r$.  In general the dual
Coxeter number, $h$, is greater than the rank, $r$.
For $SU(N)$ and $Sp(N)$ we have $h=r+1$,
and thus the abelian computation suffices to determine $W_R$
completely.  For other groups one has to extend the computation of
\DijkgraafXD\ to certain non-abelian configurations of the background
gauge fields in order to obtain the other monomials $S^k$ for $
r<k<h$.  We will content ourselves in this paper with the abelian
configurations, leaving the non-abelian configurations for future
work.  In the context of the classical groups we give a proposal of
how to extend this superpotential computation not only to the full
$W_R$ but to all powers of $S$.

\newsec{Computing the Fractional Instanton Part of the Glueball
Superpotential}

In this section we show, following \DijkgraafXD, how to compute the
glueball superpotential up to the power $S^r$, where $r$ is the rank
of the group, for arbitrary ${\cal N}=1$ supersymmetric gauge theories
in four dimensions\foot{There seems to be some confusion in the
literature on the meaning of the computation in \DijkgraafXD: This
computation is also non-perturbatively exact. In particular the
computation of the path integral contribution to glueball
superpotential reduces to perturbative configurations of charged
fields, as is clear from the derivation of \DijkgraafXD.}.  Even though
we state this in the context of ${\cal N}=1$ theories in four
dimensions, as noted in the last paper in
\dv\ this
can also be extended to ${\cal N}=1$ theories coming from higher
dimensions.

We consider turning on a constant, abelian gluino background.  This
leads to a particularly simple perturbation theory in which there are
no path ordered exponentials.  As noted in \DijkgraafXD, the
propagators of the charged fields, in the Schwinger formulation are
given by
\eqn\prop{\int \prod_i ds_i {\rm exp}\big ( -s_i \, [\,p_i^2 ~+~ \vec
{\cal W}_\alpha \cdot \vec \clam_i ~ \pi_i^\alpha ~+~ m_i\,] \big) \,,}
where $i$ denotes the edge, $p_i$ denotes the four-dimensional bosonic
momentum, $\pi_i$ denotes the fermionic momentum, and $\vec \clam_i$
denotes the charge under the Cartan generators (\ie\ the group theory
weight) flowing along the $i^{\rm th}$ edge.
Note that the bosonic and fermionic momenta,
$p_i$, $\pi_i$,
and the Cartan charges, $\clam_i$, are all conserved, and in
particular they are conserved by each vertex in a Feynman graph.  We
can therefore encode the independent variables by passing to the
corresponding loop quantities: $p_a,\pi_a,\clam_a$, where
$a=1,...,\ell$ and $\ell$ is the number of loops of the Feynman graph.

Consider a particular graph, $\cG_\ell$, with $\ell$ loops.  Introduce
the incidence matrix, $L_{ia}$,  where $L_{ia}=0$ if the $i$-th
edge does not belong to the loop $a$ and it
is $\pm 1$ (taking into account their relative orientation) if the edge
is part of the loop.  One can then write:
$$p_i=\sum_a p_a L_{ia}\,, \quad   \pi_i=\sum_a \pi_a L_{ia}\,, \quad
\clam_i=\sum_a \clam_a L_{ia}.$$
Defining an $\ell \times \ell$ matrix
$$M_{ab}(s)=\sum_i s_i L_{ia}L_{ib} \, $$
the propagators can be written as
$$\int \prod_i ds_i e^{-s_im_i}{\rm exp}\big(-[\, p_a \, M_{ab } \, p_b
~+~
\vec {\cal W}_\alpha \cdot \vec \clam_a \, M_{ab}\, \pi_b^\alpha ]
\big).$$
Integrating out the $p_a$ and $\pi_a$ leads to (up to factors of
$2\pi$):
$$\eqalign{\int \prod_i ds_i e^{-s_im_i} & det(M)^{-2}
\epsilon_{a_1...a_\ell}\, \epsilon_{b_1\dots b_\ell} \,
\big(M_{1 a_1}M_{2 a_2} \dots M_{l a_l}\big)\big( M_{1 b_1}
M_{2 b_2} \dots M_{l b_\ell}\big)\cr
\times
&\big(
   {\cal W}^{\mu_1}_1\,{\cal W}^{\mu_2}_1...{\cal W}^{\mu_\ell}_1
\big)\;\big(
{\cal W}^{\nu_1}_2  \dots {\cal W}^{\nu_\ell}_2\big)\;\big(
\clam^{\mu_1}_1
\,\clam^{\mu_2}_2 \dots  \clam^{\mu_\ell}_\ell\big)\;\big(
\clam^{\nu_1}_1\dots \clam^{\nu_\ell}_\ell\big)}$$
where the $\mu_i, \nu_j$ are vector indices on the Cartan subalgebra.
The $\epsilon$'s and the combination of $M$'s generate a $ det (M)^2$
which cancels the $det(M)^{-2}$ coming from the bosonic momentum
integration.  Thus, the $s_i$ dependence of the amplitudes trivializes
just as in the computation in \DijkgraafXD.  The $s_i$ integration can now
be performed trivially leading to $1/m_i$ for each propagator.  The
complete amplitude is then obtained by combining this with the
combinatorial factor, $F_\clam$, coming from the interaction vertices
and the symmetry factors of the graph, with fixed internal charges
$\clam_a$, which leads to the matrix model amplitude
$F(\clam )$.  We are thus left with:
$$F(\clam) \; \big(
   {\cal W}^{\mu_1}_1{\cal W}^{\mu_2}_1...{\cal W}^{\mu_\ell}_1
\big)\;\big(
{\cal W}^{\nu_1}_2  \dots {\cal W}^{\nu_\ell}_2\big)\;\big(
\clam^{\mu_1}_1
   \clam^{\mu_2}_2 \dots  \clam^{\mu_\ell}_\ell\big)\;\big(
   \clam^{\nu_1}_1\dots \clam^{\nu_\ell}_\ell\big) \,.$$
The factor $F(\clam)$ is the corresponding matrix model amplitude
with loop weights specified by $\clam$.
This needs to be summed over all the weights in the representations
running around the loops, and over all the ways the representations
that can run
through the graph, $\cG_\ell$.
Thus, associated to each such graph, $\cG_\ell$, there
is a group theory factor
\eqn\grouptensor{T^{\mu_1 \dots \mu_\ell\, \nu_1 \dots \nu_\ell}
~\equiv~
{1\over (\ell!)^2}  \, \sum_{ \clam_1, \dots,  \clam_\ell  }\,
F(\clam)\,
   \clam^{[\mu_1}_1
\clam^{\mu_2}_2 \dots \clam^{\mu_\ell]}_\ell \
   \clam^{[\nu_1}_1\clam^{\nu_2}_2 \dots \clam^{\nu_\ell]}_\ell\,,}
where the $\mu$'s and $\nu$'s are totally antisymmetrized among
themselves. The tensor is symmetric under any exchange of members in a
pair $(\mu_k, \nu_k)$, but
the very important point is that this tensor is invariant under the
Weyl group of the underlying gauge group. One can then show
that for a simple group one has:
\eqn\Tform{T^{\mu_1 \dots \mu_\ell\, \nu_1 \dots \nu_\ell} ~=~
C\, (\delta^{\mu_1\, \nu_1} \,\delta^{\mu_2\, \nu_2} \dots
\delta^{\mu_\ell\,
\nu_\ell} ~\pm~
{\rm permutations}) \,,}
for some constant $C$ \foot{Throughout this discussion we are
taking the metric on the Cartan subalgebra
to be $g_{\mu \nu}=\delta_{\mu\nu}$.}.  A proof of this statement
is given in Appendix A.
Contracting $T$ with the ${\cal W}$'s and using the definition
of the glueball field we get a contribution of $C~ \ell! ~S^\ell $ to
the glueball superpotential. To find $C$, one can contract both
\grouptensor\ and \Tform\ with
$\delta_{\mu_1\, \nu_1} \,\delta_{\mu_2\, \nu_2} \dots
\delta_{\mu_\ell\,
   \nu_\ell}$.
The left-hand side of \Tform\ gives $C ~ r!/(r-\ell)!$ and from
\grouptensor\ this is
$${1\over \ell!} \,   \sum_{ \clam_1, \dots  \clam_\ell  }\,
F(\clam)\, \det (\vec \clam_a \cdot \vec \clam_b), $$
where $\vec \clam_a\cdot \vec \clam_b$ denotes the $\ell \times \ell$
matrix
of inner products of weights in each loop.

Putting all this together, we see that the graph
$\cG_\ell$ gives a contribution to the superpotential of:
\eqn\graphcont{\big( \Delta W \big)_{\cG_\ell} ~=~  \sum_{ \clam_1,
\dots,
    \clam_\ell  }\,
F(\clam)\,    \det (\vec \clam_a \cdot \vec \clam_b) \, {(r- \ell
)!\over r!}
   \, S^\ell
   \,.}
This must then be summed over all graphs in the usual loop expansion to
obtain  $W_{R}$.

As a special case, consider the results of
\DijkgraafXD\ for $U(N)$ with adjoint matter
fields.  The determinant factor in \graphcont\ implies that the $\ell$
Cartan charges $\clam_a$ should be linearly independent.  This means
that the corresponding `t Hooft diagrams contributing to $F(\clam)$
should be planar (having fewer `t Hooft index loops is equivalent to a
linear relation between the $\clam_a$).  With no loss of generality we
can thus restrict to planar diagrams with $\ell+1$ distinct choices
for the `t Hooft index loops.  For each such choice the determinant in
\graphcont\ is the same as the determinant for the Cartan matrix of
$SU(\ell+ 1)$, which is $\ell +1$.  In \graphcont\ this is to be
summed over all the roots of the adjoint of $SU(N)$, and the number of
such index loop choices is:
$$N(N-1)...(N - \ell )={N!\over (N-\ell-1)!}\,,$$
Substituting these into \graphcont\ (with $r=N-1$) we get:

$${N!\over (N-\ell-1)!}\,(\ell+1)\,
{(N-1-\ell)!\over (N-1)!}\, S^\ell=N \,(\ell+1) \,
S^\ell $$
and there are also the combinatorial factors $F_{0,h}$
of the planar graphs with $h=l+1$ holes.

In particular if we define $F_0(S)=\sum F_{0,h}S^{h}$ then this gives
rise to the $W=NdF_0/dS$ as expected.  Note, however, that this makes
sense even if $\ell \geq N$.  For $\ell \geq N$ computation of
\graphcont\ is ambiguous: On the one hand the determinant in
\graphcont\ vanishes for $\ell \geq N$; on the other hand
$(r-\ell)!=(N-1-\ell)!$ if analytically continued is infinite. Thus
naively defining the $U(N)$ theory to be part of a sequence of
theories indexed by $N$ gives a way to regularize this computation.
This was the proposal of \dv\ motivated from some cases which was
realized in string theory.  Note that we can define a similar
completion for all the classical groups in the same way. Namely,
for a given theory for $(A,B,C,D)$ groups, we can view it as a
sequence of theories indexed by $N$. Taking the large $N$ limit we can
compute the superpotential to arbitrary high powers of $S$, and in the end
substitute in the coefficient of the monomials the finite value of $N$
in the analytic expressions. This will give a potential F-complete
definition of the theory. In the next section we discuss one meaning
for this prescription\foot{An interesting test of this for $SO(N)$
can be done as follows: We have computed the correction to glueball
superpotential only up to $S^{[N/2]}$ terms, whereas we can, in
principle, compute up to $S^{N-2}$ term.  Could the extra terms be
obtained by embedding at a larger $N$ and analytically continuing the
expressions to the smaller $N$?  This passes the test for the examples
studied in \KrausJF.}.

Note that the computation above can easily be generalized when we have
more than one simple group. One  then wants to compute the superpotential
$W(S_1,...,S_d)$ for the glueball fields $S_i$ of the $d$ simple group
factors. The tensor, \grouptensor\ is then only skew on each simple
factor, and it decomposes into $\delta$'s as in \Tform\ on each
such factor.  One obtains:
\eqn\gen{ \eqalign{\big( \Delta W(S_1,...,S_d))_{\cG_\ell}  =
\sum_{\ell_1+...+\ell_d=\ell}~\sum_{\clam_1, \dots,  \clam_\ell
}~&\sum_{\omega,\omega' \in {S_\ell\over S_{{\ell}_1}\times \ldots \times S_{\ell_d}}}
\, F(\clam^1,...,\clam^d)\epsilon(\omega ) \epsilon (\omega ')\cr
& \times \prod_{i=1}^d det_{\ell_i\times
\ell_i}(\vec \clam^i_{\omega (a)}\cdot \vec \clam^i_{\omega' (b)}) {(r_i-\ell_i)!\over
r_i!}\, S_i^{\ell_i}}}
where one chooses a partition of ${\ell}$ loops into ${\ell}_i$
loops and $\epsilon(\omega ) =\pm 1$ denote the corresponding
permutation factors of the symmetric group
$ S_{\ell}\over S_{\ell_1}\times \ldots \times S_{\ell_d}$.
Note that this expression also applies when
the choice of the vacuum breaks a gauge symmetry. As discussed in
\dgkv\ the only new ingredient is that superpotential
will include contributions of the charged ghost field as well.
One can extend the foregoing computation to include the computation
of the $U(1)$ gauge coupling constants as well, as was done for the
$U(N)$ in \DijkgraafXD,  and here we leave it to the reader.


\newsec{A Resolution of Ambiguities for Classical Groups}

As noted above, the glueball superpotential is classically ambiguous
for powers of the glueball field beyond the dual Coxeter number of the
group.  In \dv\ a prescription was given, using the `t Hooft double
line notation, to obtain all powers of the glueball field.   Let us
first review how this is done.

Consider, for concreteness, a $U(N)$ gauge theory with adjoint fields.
We can instead consider the same theory, with the same matter
content, but with group $U(Nk)$.  For low enough powers of $S$, the
dependence of the glueball superpotential on $k$ is simply a
multiplicative factor of $k$.  But, in the $U(Nk)$ theory, the
glueball superpotential is unambiguous up to $S^{Nk}$.  Thus, by taking
$k$ large enough, computing the coefficient of $S^l$, and dividing
it by $k$, gives a prescription for completing the full perturbative
series.  This amounts to getting one factor of $S$ for each `t Hooft
index loop except one, and not using the classical relation to set
$S^h=0$.  It is as if we treat different index loops as giving
distinct $S$'s.

There is an alternative, and we believe more fundamental, way to
understand this, which was pointed out in the last reference in \dv :
Consider, instead of the $U(N)$ theory, an ${\cal N}=1$ supersymmetric
theory based on the non-unitary supergroup $U(N+k|k)$, with the same
matter content and superpotential.  (See also \KawaiYF\ for another
discussion on supergroups in the N=1 context.)  This is very natural
{}from the viewpoint of brane/anti-brane system \vato, where we have
added $k$ such pairs.  For this theory, the rank is $N+2k$, so the
computation of the glueball superpotential is unambiguous up to
$S^{N+2k}$.  However the coefficients will not depend on $k$ at all.
This is clear because in each `t Hooft index loop the supertrace gives
$(N+k)-k=N$.  So there is no $k$-dependence.  In particular the answer
is the same as that of $k=0$. This $k$-independence (not even a simple
multiplicative factor) of the coefficients of the glueball
superpotential suggests that we define the $k=0$ theory in terms of
this theory by taking the $k\rightarrow \infty$ limit, which yields
unambiguous answer for the superpotential to all orders.  This is
physically analogous to saying that the $k$ brane/anti-brane pairs
disappear.

We can also apply the supergroup F-completion to $SO$ and $Sp$
theories.  For these groups using the embedding into
$SO(kN)$ or $Sp(kN)$ will not work as nicely because the
superpotential contains terms both linear in $k$ ($g=0$ contribution)
and independent of $k$ ($\IR \IP^2$ contribution) \refs{\oz ,\AshokBI,
\KlemmCY}.  But the idea of the supergroup works just as well
as it does for $U(N)$: That is, we embed the theory in $SO(N+2k|k)$
and $Sp(N+k|k)$ and then the $k$-independence works exactly as it does
for $U(N)$. This way of resolving the ambiguity boils down to treating
distinct `t Hooft loops as giving rise to distinct $S$'s, which is the
same as computing it for large $N$, and then analytically continuing
the coefficients of each power of the glueball field to finite $N$.

For $\cN =1$ supersymmetric $G(N)= U(N)$, $SO(N)$ or
$Sp(N)$, with a single adjoint, the $k$ -independence can be further
justified: The non-unitary theory based on the supergroup $G(N+k|k)$
is equivalent to a unitary one \dv.  For example, for $U(N+k|k)$, one
can take block diagonal vevs for the adjoint field breaking to
$U(N+k)\times U(k)$, with ghost fields which are bifundamental.
However, because of the super-group nature of the gauge group, the
ghost fields are ordinary rather than fermionic fields.  This is the
same as the matter content of a {\it unitary} ${\cal N}=2$
supersymmetric quiver theory based on $\widehat{A}_1$.  Thus, as far
as $F$-terms are concerned, this non-unitary system is equivalent to
that of a unitary theory.\foot{Higgsing
ordinary gauge theories, e.g. $U(N)\rightarrow \prod U(N_i)$ by the
vev of an adjoint field, are also examples where a unitary theory has
F-terms which are indistinguishable from those of a non-unitary
theory.  The original, unitary theory has massive W-bosons in the
bi-fundamental representations.  This unitary theory is F-equivalent to
the gauge fixed version, which is a non-unitary $\prod _i U(N_i)$
gauge theory, which instead has bi-fundamental ghost fields, viewed as
matter fields with the wrong statistics.  These ghosts couple to the $U(N_i)$
adjoints via superpotential terms \dgkv.  Again we see that, considering
only the F-terms, it is not easy to distinguish a unitary theory
{}from a non-unitary one.}

Finally, all of the $G(N+k|k)$ theories have a Higgs branch, where
they can be Higgsed down to the original $G(N)$ theory in the IR.
Moving the vacuum to be out along this Higgs branch corresponds to
moving away the $k$ added brane-antibrane pairs.  The IR theory is
then the original $G(N)$ theory, along with an approximately decoupled
$U(k)$ gauge theory with enhanced supersymmetry.  Infinitely far along
this Higgs branch, the $U(k)$ gauge dynamics completely decouples, and
we are left with the original $G(N)$ theory -- up to the possibility
of the residual instanton effects that we will discuss and classify.
Even at finite distances along the Higgs branch, we expect that no
dynamically generated superpotential lifts the moduli space
degeneracy, i.e. the superpotential is constant along the Higgs
branch, and thus $k$ independent.  This is expected because of the
enhanced supersymmetry of the low-energy $U(k)$ theory, and because
the Higgs branch is the probe realization of the fact that the added
brane-antibrane pairs can move around freely in the geometry.  We thus
view the $G(N+k|k)$ as a particularly natural F-completion of the
theory $G(N)$.

In the remainder of this section, we will outline the F-completion of
several examples.  The reader need not get lost in all of the examples:
the main idea is apparent in the $U(N)$ with adjoint example.  
In the next section, we will further analyze these
examples, to determine in precisely which (rare) cases there are
residual instanton effects when $G(N+k|k)$ is Higgsed to $G(N)$.
In that context, the $Sp(N)$ with anti-symmetric tensor example will
also be especially interesting.

\subsec{$U(N)$ with an adjoint and general superpotential}

Consider $U(N)$ with an adjoint $\phi$ and superpotential
$W=Tr W(\phi)$, with $W(\phi)=\sum _{p=1}^{n+1} g_p\phi ^p/p$.
When the superpotential is just a mass term, we can integrate
out the adjoint to obtain pure $U(N)$ ${\cal N}=1$ Yang-Mills
in the IR; more generally, the IR gauge group is $\prod _{i=1}^n
U(N_i)$ with $\sum _i N_i=N$.

The F-completion of this theory is $U(N+k|k)$, with adjoint $\Phi$ and
superpotential $W=Str W(\Phi)$.  We gauge fix this theory to $U(N+k)\times
U(k)$, choosing a gauge where the adjoint $\Phi$ takes the form
\eqn\Phiis{\Phi = \pmatrix{\phi & 0\cr 0& \phi '},}
with $\phi$ a $U(N+k)$ adjoint and $\phi '$ a $U(k)$ adjoint.  Associated
with this gauge choice are some bi-fundamental ghost fields which, because
of the statistics of the original supergroup, transform as bi-fundamentals
of ordinary statistics.  We thus obtain the
the following {\it ordinary} $\widehat A_1$ quiver gauge theory:
\eqn\unsgi{ \matrix{&& U(N+k)& \times& U(k)\cr &\phi
&Ad&&.\cr &\phi '&.&&Ad\cr &Q_i&\fund&&\overline{\fund}\cr
&\widetilde{ Q}_i &\overline{\fund} && \fund,\cr }} with $i=1,2$.  The
superpotential is
\eqn\waduu{W=\Tr W(\phi ) - \Tr W(\phi ')+\sqrt{2}\sum _i \left(\Tr Q_i
\phi \widetilde{Q}_i - \Tr \widetilde{Q}_i \phi ' Q_i \right) ,}
with the relative sign between the $\phi$ and $\phi '$ terms coming
{}from that of the $Str$ of $U(N+k|k)$.  The cubic interaction in
\waduu\ between the ghosts $Q_i$ and $\widetilde Q_i$ and the adjoints
$\phi $ and $\phi '$ arises in the standard Faddeev-Popov procedure,
as discussed in \dgkv.  The $\sqrt{2}$, which is henceforth not
explicitly written, is the ${\cal N}=2$ value for the coefficient
and can be absorbed into the normalization of the fields, rescaling
the mass and other superpotential couplings. 

When $W=\half m\Phi^2$, we can integrate out the adjoints to obtain
\eqn\wuul{W=-{1\over m}\Tr (Q_1\widetilde Q_1 Q_2 \widetilde Q_2 -
Q_1\widetilde Q_2Q_2\widetilde Q_1).}
This is the system studied in \refs{\klebw, \klst}, corresponding to
$k$ D3's and $N$ wrapped D5's on the resolved conifold.  This theory
has duality cascades \klst, where $k$ shifts by $N$ units at a
time.  (As do the generalizations with general higher order
superpotential \refs{\ElitzurHC , \CachazoSG}.)  This can be
used to relate the $k=0$ and $k=\infty$ theories.  So
the $k$ independence of the original $U(N+k|k)$ theory, which was a
simple consequence of the relative sign in the $Str$, is related to
Seiberg duality in the gauge fixed version \unsgi.  Further, for large
$k$ the theory is almost conformal.  So resolving the ambiguities of
the glueball superpotential by embedding the $k=0$ theory in the
$k=\infty$ theory is related to fixing the UV completion by starting
{}from a perturbed conformal fixed point.

Independent of the cascade, reducing the $k$ by one unit at a time
also makes sense physically: It corresponds to going to a particular
Higgs branch of the theory (equivalent to removing brane/anti-brane
pairs with a net 3-brane charge of 1 unit).  For example we can take
$(Q_1)_c^{c'}=(\widetilde Q_2)^c_{c'}= v_c\delta _c^{c'}$, with $\phi =
\phi ' =0$.  This is a
solution of the $D$ and $F$ term equations ($c$ is a $U(N+k)$
fundamental index and $c'$ is a $U(k)$ fundamental index).  For example,
if  $v_c=v$ for $c=1,\dots, \ell$, and zero otherwise, we Higgs
$U(N+k)\times U(k)$ to $U(N+k-\ell )\times U(k-\ell )$, along with an
approximately decoupled $U(\ell )$, ${\cal N}=4$ theory (which does not
contribute to the $F$ terms), corresponding to pulling away $\ell $
units of 3-brane charge.

The $F$ terms of the $U(N+k|k)$ theory, or its gauge fixed version
\unsgi\ are independent of the location on the foregoing Higgs branch,
which is to say that they are independent of $k$.  So it is reasonable
to compute $F$ terms for the original, $k=0$, $U(N)$ theory in terms
of the large $k$ $U(N+k|k)$ theory.  However, as will be discussed in the
next section, this F completion of $U(N)$ into $U(N+k|k)$ can
differ, in some rare instances, from the more standard UV completion.
The potential differences arise at the last steps of Higgsing, for
small $k$.

Now consider the dynamical scales of the foregoing theories.  The
original $U(N)$ theory has a scale $\Lambda$, with the instanton
factor $\Lambda ^{2N}$.  We take the $U(N+k|k)$ theory to have this
same scale $\Lambda$, so that is the scale of both $U(N+k)$ and $U(k)$
upon gauge fixing.  The $U(N+k)$ theory has an adjoint and $N_f=2k$
flavors, so its instanton factor is $\Lambda _1^{2(N+k)-2k} =\Lambda
^{2N}$, which we equate with the instanton factor in the original
$U(N)$.  On the other hand, the $U(k)$ factor has $N_f=2(N+k)$, so its
instanton factor is $\Lambda ^{2k-2(N+k)}=\Lambda ^{-2N}$, which is
inverse to the original $U(N)$ instanton factor.  This relative sign
is related to that of the beta functions, coming ultimately from the
relative sign in the supertraces.

\subsec{$SO(N)$ with a single adjoint $\phi$ and superpotential $Tr
W(\phi)$.}

By considering an orientifold of the foregoing $U(N+k)\times U(k)$ theory,
one sees that the natural F-competion of $SO(N)$ is $SO(N+2k)\times Sp(k)$.
(We label $Sp$ groups by their rank, so that $Sp(1)\cong SU(2)$.)
This group is indeed the gauge-fixed version of the supergroup $SO(N+2k|k)$
\foot{We write $SO(N|M)$ for the supergroup with bosonic part $SO(N)\times
Sp(M)$.  The names $B(n,m)$ and $D(n,m)$ are sometimes used for
what we call $SO(2n+1|m)$ and $SO(2n|m)$, respectively.}.
So our $F$-completion of $SO(N)$ is $SO(N+2k|k)$ with an adjoint
$\Phi$ and superpotential  $Str W(\Phi)$. We can gauge
fix $SO(N+2k|k)$ to $SO(N+2k)\times Sp(k)$ by choosing the adjoint as
\Phiis.  This gauge choice requires Faddeev Popov ghosts which are,
as before, ordinary superfields.  Because the $SO(N+2k|k)$ adjoint
$\Phi$ satisfies a reality condition, so do the ghosts.  This implies
that, rather than two hypermultiplets, as in \unsgi, the ghosts are now
two half-hypermultiplets.  So the gauge fixed version of our F-completion
is the following {\it ordinary} gauge theory:
\eqn\sosgi{ \matrix{&& SO(N+2k)& \times& Sp(k)\cr &\phi
&Ad&&.\cr &\phi '&.&&Ad\cr &Q_i&\fund&&{\fund}\cr
}} with $i=1,2$.
The superpotential is
\eqn\wado{W=\Tr W(\phi _1) - \Tr W(\phi _2)+\sum _{i=1}^2 \left(\Tr Q _i
\phi \widetilde{Q}_i - \Tr \widetilde{Q}_i \phi ' Q_i \right),}
where we define $(\widetilde{Q}_i)^c_{c'}=\delta ^{cd}J_{c'd'}(Q_i)_d^{d'}$,
which is the reality condition for half-hypermultiplets, with $c,d$ $SO$
fundamental indices and $c'd'$ the $Sp$ fundamental indices.

When $W=\half m\Phi^2$, we can integrate out the adjoints to obtain
\eqn\wol{W=-{1\over m}\Tr (Q_1\widetilde Q_1  Q_2  \widetilde Q_2-
Q_1\widetilde Q_2 Q_2\widetilde Q_1).}
The $SO(N+2k)$ theory has $N_f=4k$
vectors and the $Sp(k)$ theory has $N_f=N+2k$ flavors.  There is a
duality cascade where $SO(N+2k)$ group gets strong and dualized
\IntriligatorID\ into
$SO(2k-N+4)$.  Then the $Sp(k)$ gets strong and is dualized
\IntriligatorNE\ into $Sp(k-N+2)$.  After these two steps, we get back
the same theory with $k\rightarrow k-N+2$.  This cascade, and its string
theory realization, was discussed in \NaculichPQ.
Exactly the same cascade occurs for theory
\sosgi\ with adjoint and general tree-level superpotential; this is seen via
a deformation of the dualities discussed in \refs{\LeighQP, \IntriligatorAX}.

In terms of orientifolding $N$ wrapped D5's and $k$ D3's on the conifold,
the gauge theory would naturally be similar to \sosgi, but with gauge
group $SO(N+2k)\times Sp(k-1)$.  (For example,  this theory is superconformal 
for $N=0$, whereas \sosgi\ is not.)  In the large $k$ limit, these two
F-competions are, of course, completely equivalent.  But we prefer
\sosgi, because it recovers the original $SO(N)$ theory for $k=0$.

Independent of the cascades, the theory \sosgi\ has a Higgs branch,
along which we can successively reduce $k\rightarrow k-1$.  E.g. we
can take $(Q_1)_c^{c'}=(\widetilde{Q}_2)^c_{c'}=v_c\delta _c^{c'}$;
taking $\ell$ equal non-zero $v_c$, we get a similar theory, with
$k\rightarrow k-\ell$, along with an approximately decoupled $U(\ell)$
${\cal N}=4$ theory.   The $F$ terms are independent of the location
of the theory along this Higgs branch.  So the large $k$ $SO(N+2k|k)$
theory indeed provides an F-completion of the original $SO(N)$ theory.

Again, we take the $SO(N+2k|k)$ theory to have the same scale, $\Lambda$,
as the original $SO(N)$ theory.  For all $k$ the $SO(N+2k)$ gauge theory
in \sosgi\ has instanton factor $\Lambda ^{2(N+2k-2)-4k}=\Lambda ^{2(N-2)}$,
which is that of the original $SO(N)$ theory.  On the other hand,
the $Sp(k)$ group in \sosgi\ has instanton factor
$\Lambda ^{2(k+1)-(N+2k)}=\Lambda ^{-(N-2)}$, which is the inverse
square-root of the original $SO(N)$ instanton factor.  This will play
a small role in the discussion in the following section.

\subsec{$Sp(N)$ with a single adjoint $\phi$ and superpotential
$W(\phi)$.}

We write the supergroup F-completion as $Sp(N+k|k)$, which denotes
the same group as $SO(2k|N+k)=D(k,N+k)$ in the above notation.  There
is an $Sp(N+k|k)$ adjoint $\Phi$ and superpotential $Str W(\Phi)$.  Upon
gauge fixing, we obtain the ordinary gauge theory
\eqn\spsgi{ \matrix{&& Sp(N+k)& \times& SO(2k)\cr &\phi
&Ad&&.\cr &\phi '&.&&Ad\cr &Q_i&\fund&&{\fund}\cr
}} with $i=1,2$.
The superpotential is
\eqn\wadp{W=\Tr W(\phi ) - \Tr W(\phi ')+\sum _{i=1}^2\left(\Tr Q_i
\phi  \widetilde{Q}_i - \Tr \widetilde{Q}_i \phi ' Q_i \right) .}
As in the previous subsection, the fields $Q_i$, which arise as the
ghosts in the gauge fixing, satisfy a half-hypermultiplet reality condition,
$\widetilde Q^c_{c'}=J^{cd}\delta _{c'd'}Q_d^{d'}$, where $c,d$ are $Sp$
fundamental indices and $c'd'$ are $SO$ fundamental indices.
When $W=\half m\Phi^2$, we can integrate out the adjoints to obtain
the quartic superpotential \wol.

As mentioned in the previous section, in an orientifold the group
would actually naturally be $Sp(N+k)\times SO(2k+2)$.  Again, this would
be equivalent to \spsgi\ in the large $k$ limit, but we prefer \spsgi\ because
it reduces to our original gauge theory for $k=0$.

The foregoing theory, with any $W(\phi)$, undergoes a RG cascade. First, the
$Sp(N+k)$ gets strong and is dualized to $Sp(k-N-2)$; next, the
$SO(2k)$ gets strong and is dualized to $SO(2k-4N-4)$.  After these
two steps, we are back to a similar theory, with $k\rightarrow k-2N-2$.
This cascade can be regarded as the continuation, to negative $N$, of
that of the previous section.

Again, independent of the cascade, and for all $W(\phi)$, there is a
Higgs branch where we can successively reduce $k\rightarrow k-1$.  The
$F$ terms are independent of the location of the vacuum on this Higgs
branch moduli space of vacua.  So the large $k$ $Sp(N+k|k)$ theory is
indeed an F-completion of the original $Sp(k)$ theory.

We take the $Sp(N+k|k)$ theory to have the same scale, $\Lambda$,
as the original $Sp(N)$ theory.  The $Sp(N+k)$ theory has
instanton factor $\Lambda
_1^{2(N+k+1)-2k} =\Lambda ^{2(N+1)}$, which is that of the original
$Sp(N)$ theory.  The $SO(2k)$ has $4(N+k)$ vectors, so its instanton factor is
$\Lambda _2^{2(2k-2)-4(N+k)}=\Lambda ^{-4(N+1)}$, which is the inverse
square of the $Sp(N)$ instanton factor.

\subsec{$SO(N)$ with a symmetric tensor $\phi$ and superpotential $W(\phi )$}

The F-completion is again $SO(N+2k|k)$, with a two-index tensor with
$\Phi =\Phi ^T$, as opposed to $\Phi =-\Phi ^T$ for the adjoint
(transpose is defined with the usual sign in the upper block and the
opposite in the lower), and superpotential $W= Str W(\Phi)$.  Gauge
fixing as in \Phiis\ yields the ordinary gauge theory \eqn\sosgi{
\matrix{&& SO(N+2k)& \times& Sp(k)\cr &\phi &\sym &&.\cr &\phi '&.&&\anti \cr
&Q_i&\fund&&{\fund}\cr}} with $i=1,2$.  The gauge fixing procedure
yields the tree-level superpotential
\eqn\wsoop{W=\Tr W(\phi ) - \Tr W(\phi 'J)+\epsilon ^{ij}\left(
\Tr \widetilde Q_i\phi Q_j+\Tr Q_i \phi ' \widetilde Q_j\right),}
with repeated indices summed and $\widetilde Q_i \equiv Q_i ^TJ$, and   
$J^{[c'd']}$ is the $Sp(k)$ defining symplectic tensor. 

When the superpotential is a mass term, we can integrate out $\phi$
and $\phi'$ to obtain
\eqn\wpal{W=-{1\over m}\Tr (Q_1\widetilde Q_1 Q_2 \widetilde Q_2
-Q_1\widetilde Q_2 Q_2\widetilde Q_1).}
For the general theory with superpotential $W(\Phi)=Str \Phi ^{p+1}$ there
is a duality cascade, which can be obtained by a deformation
(corresponding to the cubic interactions in \wsoop) of that
discussed in \refs{\IntriligatorFF, \IntriligatorAX}.  Unlike the above
cases, the groups in the cascade now depend on $p$.  First the $SO$ gets
strong and is dualized to $SO(2k-N+4p)$, then the $Sp$ group gets
strong and is dualized to $Sp(k-N+2p)$; after these two steps we are
back to a theory which is similar to the original theory, but with
$k\rightarrow k-N+2p$.

Again, independent of the cascade reduction of $k$, there is a Higgs
branch, along which we can successively reduce $k\rightarrow k-1$.
For example, taking $\ev{Q_1}=diag(v_1, v_1, v_2, v_2, \dots v_l, v_l,
0, 0 \dots )$, with all other entries zero and with
$\ev{Q_2}=\ev{\phi} =\ev{\phi '}=0$ solves the $D$ and $F$ term
equations.  Along this Higgs branch we reduce to a similar theory,
with $k\rightarrow k-\ell$, together with an approximately decoupled
${\cal N}=2$ $U(\ell)$ theory.  The exact $F$ terms are again expected
to be independent of the location of the vacuum on this Higgs branch;
showing that the large $k$ $SO(N+2k|k)$ theory is indeed a sensible
F-completion of the $k=0$ theory.

We take $SO(N+2k|k)$ to have the same scale $\Lambda$ as the
original $SO(N)$ theory.  For all $k$, the $SO(N+2k)$ gauge
group has instanton factor $\Lambda ^{3(N+2k-2)-(N+2k+2)-4k}=
\Lambda ^{2N-8}$, which is the instanton factor in the original
$SO(N)$.   The $Sp(k)$ has instanton factor
$\Lambda ^{3(k+1)-(k-1)-N-2k}=\Lambda ^{4-N}$, which is the inverse
square-root of the $SO(N)$ instanton factor.

\subsec{$Sp(N)$ with an antisymmetric tensor $\phi$ 
and superpotential $W(\phi)$.}

This can be regarded as the being the continuation of the previous
example to negative $N$.  The F-completion is $Sp(N+k|k)$,
and gauge fixing yields the ordinary gauge theory
\eqn\spagi{ \matrix{&& Sp(N+k)& \times& SO(2k)\cr &\phi 
&\anti &&.\cr &\phi '&.&&\sym\cr &Q_i&\fund&&{\fund}\cr}}
$i=1,2$, with superpotential
\eqn\wspuu{W=\Tr W(\phi J) - \Tr W(\phi ')+ \epsilon ^{ij}\left( 
\Tr \widetilde Q_i \phi Q_j +\Tr Q_i \phi ' \widetilde Q_j\right),}
with $\widetilde Q_i \equiv Q_i^T J$, and $J^{cd}$ is the $Sp(N+k)$
symplectic tensor.  

Again, this theory has a RG cascade where, after two steps, we return
to the same theory with $k\rightarrow k-2N-2p$ for superpotential $W(\Phi)
=Str \Phi ^{p+1}$.  Also, independent of the cascade, there is a Higgs branch
where we can successively reduce $k\rightarrow k-1$.  An example
 direction in this Higgs branch of vacua is 
$\ev{Q_1}=diag(v_1, v_1, v_2, v_2, \dots , v_\ell , v_\ell , 0,0...)$, with all
other entries zero and with $\ev{Q_2}=0$ and $\ev{\phi}=\ev{\phi '}=0$.
Along this direction we Higgs to a similar theory, with $k\rightarrow k-\ell$,
along with an approximately decoupled ${\cal N}=2$
$U(\ell)$ theory, having Coulomb branch moduli $(v_1, \dots v_\ell)$.  
 The $F$ terms are expected to be independent of the location of the
vacuum along this Higgs branch.

The $Sp(N+k)$ has instanton factor $\Lambda ^{3(N+k+1)-(N+k-1)-2k}=
\Lambda ^{2N+4}$, which is identified with the instanton factor
for the original $Sp(N)$ theory with anti-symmetric tensor.  The
$SO(2k)$ has instanton factor $\Lambda ^{3(2k-2)-(2k+2)-4(N+k)}=
\Lambda ^{-4N-8}$, which is the inverse square of the original $Sp(N)$
instanton factor.

\subsec{$U(N)$ ${\cal N}=1^*$}

The F-completion of the $U(N)$ ${\cal N}=1^*$ theory is a $U(N+k|k)$ with
three adjoints $\Phi _I$, $I=1,2,3$, and superpotential
\eqn\wstar{W=Str[ \sum _{IJK}(\Phi _I[\Phi _J,\Phi _K])\epsilon ^{IJK}+ \half m
\sum _I\Phi _I^2].}
We write the $\Phi _I$ as:
\eqn\phiix{\Phi _i =\pmatrix{\phi _i &R_i\cr
\widetilde{R}_i &\phi '_i}, \quad i=1,2, \qquad \Phi _3=\pmatrix{\phi _3
&0\cr 0&\phi _3'},}
where the zeros in $\Phi _3$ are our gauge choice in breaking $U(N+k|k)$
to $U(N+k)\times U(k)$.  The bi-fundamentals $R_i$ and $\widetilde R_i$
are fermionic.  Also, as before, the Faddeev-Popov ghosts are ordinary
matter bi-fundamentals $Q_i$ and $\widetilde Q_i$; so the gauge fixed
gauge group and matter content is the non-unitary theory (because of
the $R_i$ and $\widetilde R_i$):
\eqn\unssgi{ \matrix{&& U(N+k)& \times& U(k)\cr &\phi _I
&Ad&&.\cr &\phi '_I&.&&Ad\cr &Q_i&\fund&&\overline{\fund}\cr
&\widetilde{ Q}_i &\overline{\fund} && \fund,\cr
&R_i&\fund&&\overline{\fund}\cr &\widetilde{R}_i &\overline{\fund} &&
\fund,\cr}} with $I=1,2,3$ and $i=1,2$. The superpotential \wstar\
yields
\eqn\waduuu{\eqalign{W&=\sum _{IJK}
\epsilon ^{IJK}[\Tr(\phi _I[\phi _J,\phi _K])-
\Tr (\phi '_I[\phi '_J,\phi ' _K])]+
\sum _I{m\over 2}(\Tr \phi _I^2-\Tr \phi _I'^2)+\cr
&+\sum _i m R_i \widetilde R_i +\Tr \phi _3(Q_i\widetilde Q_i 
+R_1\widetilde{R}_2-R_2\widetilde{R}_1)-\Tr \phi
_3'(\widetilde{Q}_iQ_i+ \widetilde{R}_1R_2-\widetilde{R}_2R_1).}}  The
fermionic fields $R_i$ and $\widetilde R_i$ are massive, as are the
adjoints $\phi _I $ and $\phi '_I$.  The massless spectrum consists of
just the bi-fundamentals $Q_i$ and $\widetilde Q_i$ with a quartic
superpotential.  This is the same massless spectrum as for the
F-completion of ${\cal N}=1$ super-Yang-Mills, as expected.  We need
to keep the effects of the massive fields in \unssgi\ to see the
difference between ${\cal N}=1^*$ and ${\cal N}=1$.

Once again, the full $U(N+k|k)$ F-completion of ${\cal N}=1^*$ has
a Higgs branch, along which we effectively reduce $k$ one unit at a time.
Again, the superpotential is expected to be independent of the
location of the vacuum along this Higgs branch, implying that the
different Higgs branches, with different
values of $k$, all have the same superpotential.

\subsec{Classical groups with {\it arbitrary
massive representations}.}  

We will assume that there are no baryon operators in the
superpotential; if there is one can presumably treat it using ideas
similar to \bmrt.  We will use analytic continuation as a
regularization scheme: Compute the coefficient of $S^k$ for fixed $k$
as a function of $N$ for $k<N$.  This computation is unambiguous; one
can then analytically continue the coefficient to $k>N$.  For
representations obtained from tensor products of fundamental
representations ({\it e.g.}, not the spinorial representations of the
$SO$ groups) this is equivalent to considering multi-line Feynman
graphs and putting only up to two gluino fields per index loop and
treating each index loop independently of the other.  Again, a
justification/meaning of this regularization scheme is to embed the
theory in the corresponding supergroup, $G(N+k|k)$ and take
$k\rightarrow \infty$ to remove the ambiguities.  This prescription is
equivalent to defining the $G(N)$ theory as a Higgs branch of the
arbitrarily large rank group $G(N+k|k)$.

\subsec{Prescription for Non-Classical Groups $(E,F,G)$.}

For non-classical groups $(E,F,G)$ there is no obvious canonical
choice for the resolution of the ambiguities in the glueball
superpotential. In some cases, with certain matter content, there are
phases where the gauge group breaks to subgroups involving only
classical groups, in which case we can write the theory in terms of
the broken factors and use the above prescription to remove the
ambiguities.  For example, this is true for the $G_2$ example studied
in \refs{\gid,\rosz}, where there is a phase where $G_2$ breaks to
$SU(2)$).  In some other cases ``analytic continuation'' in groups
suggests an answer as we will see in the next section where we discuss
the ${\cal N}=1^*$ theory for arbitrary simply-laced groups.

\newsec{Comparision with Other UV Completions}

As seen in the previous section, $G(N+k|k)$ is a sensible F-completion
of $G(N)$ because the $G(N+k|k)$ theory always has a Higgs
branch\foot{ If necessary one can always add a very massive adjoint to
achieve this, without affecting the IR dynamics of the theory.}, where
we can successively reduce $k\rightarrow k-1$.  In this way, we can
Higgs from $G(N+k|k)$ back down to $G(N)$, and the exact
superpotential is expected to be independent of the location of the
vacuum on this Higgs branch.  However, as we discuss in this section,
the F-terms obtained in this way for the $G(N)$ theory can differ from
that of the standard UV completion of $G(N)$.  In terms of our
Higgsing from $G(N+k|k)$ to $G(N)$, this is seen as residual instanton
effects, associated with the broken part of the group,
$G(N+k|k)/G(N)$.

Suppose that a certain theory has gauge group $\cG$ in the extreme
UV, but the group is Higgsed at some high scale to a subgroup
$\cH$ in the IR.  Usually, at least when the Higgsing scale is taken
to the extreme UV, the IR dynamics can be understood
purely in terms of that of the IR theory $\cH$, without having to include 
additional interactions, which require knowing about
the original UV completion $\cG$.  We refer to this as ``decoupling''.
For example, for $SU(N_c)$ SQCD with $N_f<N_c-1$, the 
superpotential \AffleckMK\ for vacua far out along the classical moduli
space (as well as exactly) simply comes from gaugino condensation in
the low energy theory with $\cH = SU(N_c-N_f)$.  One known exception
to this decoupling intuition are the Wess-Zumino terms, which can be
necessary to add to the IR theory to account for any discrepancies between
the symmetries or anomalies of the UV and IR theories.  
For F-terms, there is only one known exception to the decoupling
intuition: instantons in the broken part, 
$\cG/\cH$, of the UV group $\cG$.  For example, this is how the
superpotential arises for SQCD with $N_f=N_c-1$, where the IR group
$\cH$ is trivial.  Another example is ${\cal N}=2$ SYM, where $\cH$
is the Cartan subalgebra, and instantons in $\cG/\cH$ are essential
to understanding the IR dynamics of the Coulomb branch \SeibergRS.

Instantons in partially broken UV groups can appear to violate
decoupling intuition, since the particular UV completion can affect
the IR results.  For example, $\cN =1$, $\cH=SU(2)$, super-Yang-Mills
has a UV completion, based on $\cG =SU(2)\times SU(2)$ which is broken
to $\cH$ at an arbitrarily high scale in the UV, where instanton
effects in the broken UV group $\cG /\cH$ are of the same size as the
leading non-perturbative effects in the IR $SU(2)$, and can cancel the
IR $\cH$ gaugino condensation superpotential \refs{\IntriligatorJR,
\IntriligatorID}.  (In this case, the instantons in the broken
$\cG/\cH$ actually compete with the fractional instantons in $\cH$.
This is unusual, since generally $\cG /\cH$ instantons are similar to
whole $\cH$ instantons; here it's fractional because of how the IR
group $\cH$ is diagonally embedded in the UV group $\cG$.)

When can such instantons in the partially broken group $\cG/\cH$
potentially contribute?  A standard lore about instantons, which is
borne out in the exact results for supersymmetric gauge theories, is
that they never contribute to the effective action unless the gauge
group is sufficiently Higgsed\foot{ Technically, instantons don't
exist in this case -- one has to consider constrained instantons.}.
The Higgsing is needed to regulate the divergence in the integral over
instanton size\foot{Alternatively, this divergence can be regulated by
considering correlation functions at separated points, in which case
no Higgsing is needed.  This is how instantons contribute to the
glueball $h$-point function $\ev{S^h}$ in supersymmetric Yang-Mills,
where there is no Higgsing of the gauge group.  We will only consider
superpotential terms, in which case the sufficient Higgsing is
required for an instanton contribution.}.  What ``sufficiently Higgsed''
means is that $\cG$ instantons can contribute only if the Higgsed
part $\cG/\cH$ has $\pi _3(\cG/\cH)\neq 0$.  See \CsakiVV\ for a nice
discussion.

Applying these ideas to our F-completion, we have $\cG=G(N+k|k)$ and
$\cH = G(N)$.  Our F-completion $G(N+k|k)$ of $G(N)$ can include
residual instanton F-term contributions, coming from $G(N+k|k)$, which
are not present in the standard UV completion of the $G(N)$ theory.
Such residual instanton effects can arise when $\pi
_3(G(N+k|k)/G(N))\neq 0$.  As we will discuss, this is a necessary,
but not sufficient, condition for such effects.  In a nutshell, the
$\pi _3(\cG/\cH) \neq 0$ condition here implies that $\cG$ has to have some
$SU(2)$ factor, which is either completely broken, or broken to
$U(1)$, in $\cH$.  The additional condition is on the number of
massless flavors in this broken $SU(2)$.  The usual condition for an
$SU(2)$ instanton contribution to the superpotential, based on a
zero-mode analysis, is that the $SU(2)$ must have precisely $N_f=1$
flavor.  For our examples, because of a tree-level superpotential, the
relevant number of $SU(2)$ flavors for an instanton contribution is
instead $N_f=2$ or $N_f=3$ massless flavors.  These are the only two
possibilities; for example, there is no superpotential contribution for
$N_f=4$ massless flavors.

To give an example of the kind of effect which these residual
instanton contributions from $G(N+k|k)$ have, consider the
glueball superpotential for pure supersymmetric Yang-Mills:
\eqn\wsymb{W=hS\left(\log\left({S\over \Lambda
^3}\right)-1\right).}
The standard gauge theory (``sgt'') glueball
superpotential for the classical gauge groups is given by \wsymb\ with
\eqn\hsgt{\eqalign{U(N):\qquad h_{sgt}&=N-\delta _{N,1}\cr
Sp(N):\qquad h_{sgt}&=N+1-\delta _{N,0}\cr SO(N):\qquad
h_{sgt}&=N-2+\delta _{N,1}+2\delta _{N,0},}} 
where the $\delta _{N,*}$ are some exceptions, corresponding, 
for example, to the fact that standard
$U(1)$ gauge theory does not have any non-perturbative effects which
could lead to photino condensation and an associated non-zero
superpotential.  There are similar exceptions  for the trivial groups 
$Sp(0)$, $SO(1)$ and $SO(0)$.

On the other hand, our $G(N+k|k)$ F-completion can not have any such
exceptions, for any $N$ values, since we can avoid the low $N$
exceptions by making $k$ sufficiently large.  The $N$ dependence
of the $G(N+k|k)$ F-completion must be completely smooth!   For example,
the F-completion for pure Yang-Mills gives the glueball superpotential
\wsymb\ with
\eqn\wlowuiv{\matrix{&U(N):\qquad &h=N\ \ \ \ \ \cr
&Sp(N):\qquad &h=N+1\cr &SO(N):\qquad &h=N-2,}} 
for all $N\geq 0$,
with no exceptions.  In particular, we find non-trivial glueball
superpotentials for $U(1)$, $Sp(0)$, $SO(1)$, and $SO(0)$.  As we will
discuss, the difference between \hsgt\ and \wlowuiv\ can be understood
as residual instanton effects, with the effect of the residual
instantons being precisely that needed to smooth out the exceptions.
For example, the glueball superpotential for $U(1)$ can be seen in its
$U(1+k|k)$ F-completion, which arises as a residual instanton effect
when we Higgs $U(2|1)$ to $U(1)$.  Another UV completion of $U(1)$
gauge theory that leads to the same non-zero superpotential of
\wsymb, with $h=1$, is non-commutivity.  For this, the
instanton leading to the non-zero $W$ can be seen directly in the
non-commutative $U(1)$ gauge theory.

We now discuss various examples.  The main idea is illustrated in the
$U(N)$ case, so the reader can feel free to skip the later
subsections.  One highlight, though, is the case of $Sp(N)$ with an
anti-symmetric tensor, discussed in sect. 5.4.  For that theory, the
residual instanton effects of our F-completion play an especially
prominent role.

\subsec{The $U(N+k|k)$ F-completion of $U(N)$ with an adjoint}

The $U(N+k|k)$ F-completion was discussed in section 4.1.  We first
consider  pure ${\cal N}=1$ $U(N)$ Yang-Mills, which we get
{}from the theory with an added adjoint by taking the superpotential to be
a large mass term.  The low energy theory is then the $\widehat A_1$
quiver $U(N+k)\times U(k)$ theory, with no adjoints, and with a quartic
low-energy superpotential \wuul, which we write as
\eqn\wuull{W=-{1\over m}\Tr (M_{11}M_{22}-M_{12}M_{21}),}
with $(M_{ij})_a^b=(Q_i)_a^{b'}(\widetilde{Q}_j)_{b'}^b$.

When we Higgs $k\rightarrow k-1$, there can only be a residual
instanton effect if $\pi _3(U(N+k)/U(N+k-1))\neq 0$ or $\pi _3(
U(k)/U(k-1))\neq 0$.  This only happens when $N+k=2$ or when $k=2$,
i.e. when a $U(2)$ factor is Higgsed to
$U(1)$.  Even for these cases, there is generally no residual instanton
effect: if the $U(2)$ has $N_f>3$ massless flavors, the classical moduli
space is unmodified and there is a supersymmetric vaccum of the theory
with superpotential \wuull\ with zero contribution to the superpotential.
There is only the possibility of a residual instanton effect when 
$N_f\leq 3$.

Noting that the $U(N+k)$ has $N_f=2k$ and the $U(k)$ has $N_f=2(N+k)$,
we see that for $k=2$ one has $N_f=2N+4$ massless flavors, which is
too many for a residual instanton contribution for any $N\geq 0$.  So
the only possibility for a residual instanton contribution is in the
$U(N+k)$ factor, which requires $N+k=2$ and $0<2k<4$, \ie\ $N=k=1$.
To summarize, there is only the possibility of a residual instanton
effect for $U(N)$ and that is when  $N=1$, where the residual instanton can
only contribute at the very last stage in the $k\rightarrow k-1$ Higgsing
of $U(N+k|k)$, namely $U(2|1)\rightarrow U(1)$.

So we consider the $U(2)
\times U(1)$ theory with $N_f=2$ bi-fundamentals and tree level superpotential
\wuull.  Since $U(2)$ has $N_f=2$ flavors there is an instanton
effect, which leads to a quantum modified moduli space constraint
\Seiberg, $\det M-B\widetilde B = \Lambda _L^4$, where $\Lambda _L$ is
the scale of the low energy $U(2)$ gauge theory, in which the adjoint
has been integrated out.  The matching relation to the scale $\Lambda$
of the high energy $U(2)$ theory which includes the adjoint of mass $m$ is
$\Lambda _L^4=m^2\Lambda ^2$.  We write the superpotential as
\eqn\wsuiis{W=S\log \left({\det M - B\widetilde B\over
m^2\Lambda ^2}\right)-{1\over m}\det M,}
where the glueball field $S$ term is to enforce the quantum moduli
space constraint, and the other term is \wuul.  

We now integrate out
the fields $M$ and $B$ by their equations of motion
\eqn\mbeom{S M^{-1}= {1\over m} M^{-1}\det M, \qquad SB=S\widetilde B
=0.}
There are two branches of vacua.  One is an 
isolated vacuum at $\ev{M}=0$, $\ev{S}=0$,
and $\ev{B\widetilde B}=-\Lambda ^4$. This is the wrong solution
for our purposes, because it does not have the Higgs branch connecting to
our original $U(1)$ theory in the IR.  The relevant solution, which does
have the expected Higgs branch, is 
\eqn\higgsm{ \ev{M_{11}M_{22}-M_{12}M_{21}}=m\ev{S}, \qquad\hbox{and}
\qquad  \ev{B}=\ev{\widetilde B}=0.}
The $\ev{M_{ij}}$ of \higgsm\ is a three complex dimensional moduli
space of vacua, which is the deformed conifold geometry, with
$m\ev{S}$ giving the deformation.  This is the expected moduli space
of the added D3 brane, probing the geometry of the string theory realization
of this theory.  
Everywhere on this moduli space, even infinitely far from the
origin, where $U(2)\times U(1)$ is Higgsed to our original $U(1)$ theory
(along with an approximately decoupled, ${\cal N}=4$, $U(1)$ theory)
in the extreme UV, there
is the constant superpotential for the glueball superfield
\eqn\wlowii{W=S\left(\log\left({S\over \Lambda _{U(1)}^3}\right)-1\right),}
where $\Lambda _{U(1)}^3=m \Lambda ^2$ is naturally regarded as the
scale of the low energy $U(1)$ pure Maxwell theory, after integrating
out the massive adjoint.  Thus our F-completion of $U(1)$ has the
superpotential
\wsymb, with $h=1$.  If we integrate out $S$, we obtain 
$\ev{S}=\Lambda _{U(1)}^3$
and the constant superpotential along the above Higgs branch
\eqn\wlowiii{W_{low}=-\Lambda _{U(1)}^3,}
coming from the instanton in the broken $U(2)$ of the extreme UV completion. 

Let us compare this with $U(N+k|k)$ F-completion of $U(N)$ for $N>1$.  
Consider, in particular, the gauge fixed $U(N+k)\times U(k)$ theory \unsgi\
for $N>1$ and $k=1$.  The superpotential is
\eqn\wcompare{W=S\left[ \log \left({S^{N-1}\det _{ij}
M_{ij}\over m^{N+1}\Lambda ^{2N}}\right) -(N-1)
\right]-{1\over m}\det M,}
where $i=1,2$.  The first term in \wcompare\ is dynamically generated
by the $U(N+1)$ theory with $N_f=2$ flavors and the second is the
tree-level term \wuul.  The $M_{ij}$ equation of motion is
$SM^{-1}={1\over m} M^{-1}\det M$, as in \mbeom.  The solution is as
in \higgsm: $\ev{\det M}=mS$, giving the expected deformed conifold
moduli space of vacua, along which the gauge group can be Higgsed to
the original $U(N)$ theory (along with some decoupled additional
fields) at an arbitrarily high scale in the UV.  Along this entire
moduli space, the superpotential has the constant value:
\eqn\wcomparee{W=S\left[ \log \left({S^N\over 
m^{N}\Lambda ^{2N}}\right) -N
\right],}
which, is precisely the expected gaugino condensation superpotential
for the $U(N)$ theory obtained in the IR, out along the Higgs branch.
The IR $U(N)$ theory has scale $\Lambda _{U(N)}$ given by $\Lambda
_{U(N)}^{3N}=m^N\Lambda ^{2N}$, upon integrating out the massive
adjoint.  No residual instanton contribution is present or needed for
$N>1$, since
\wcomparee\ can be seen directly in terms of the IR $U(N)$ theory for $N>1$.

The upshot is that the residual instanton term only arises for $U(N)$
with $N=1$, and it is precisely such that it eliminates the $\delta
_{N,1}$ in
\hsgt, yielding the superpotential \wsymb\ with $h$ given by 
\wlowuiv, which is smooth for all $N\geq
0$.  We note, as a special case, that there is no residual instanton
superpotential for $N=0$, since then the UV completion is $U(k)\times
U(k)$ and, for the relevant case of Higgsing $k=2$ to $k=1$, each
$U(2)$ has $N_f=4$ massless flavors, which is too many to lead to a
residual superpotential.  This agrees with the fact that
\hsgt\ is already smooth for $N=0$.

Now consider the $U(N)$ theory with adjoint $\phi$ and higher order
superpotential $W=\Tr W(\phi)$, e.g. $W(\phi) = {g\over 3}\phi ^3
+{m\over 2}\phi ^2$.  Since $W'(z)=gz(z-m/g)$, the gauge group is
Higgsed in the various vacua as $U(N_1+N_2)\rightarrow U(N_1)\times
U(N_2)$.  Our F-completion of this is $U(N_1+N_2
+k_1+k_2|k_1+k_2)\rightarrow U(N_1+k_1|k_1)\times U(N_2+k_2|k_2)$.
Gauge fixing, the relevant IR gauge theory is a $U(N_1+k_1)\times
U(N_2+k_2)\times U(k_1)\times U(k_2)$ quiver gauge theory, with two
bi-fundamentals and two anti-bifundamentals connecting every pair of
gauge groups.  These bi-fundamentals come {}from the various ghosts,
with statistics, and masses given by
\eqn\bifundl{\matrix{&\hbox{\bf Groups connected}&\qquad \hbox{\bf
statistics}&\qquad\hbox{\bf mass}\cr
&\hbox{$U(N_1+k_1)$ and $U(k_1)$}&\qquad \hbox{ordinary}&\qquad
\hbox{massless}\cr
&\hbox{$U(N_2+k_2)$ and $U(k_2)$}&\qquad \hbox{ordinary}&\qquad
\hbox{massless}\cr
&\hbox{$U(N_1+k_1)$ and $U(k_2)$}&\qquad \hbox{ordinary}&\qquad
\hbox{$\Delta = m/g$}\cr
&\hbox{$U(N_2+k_2)$ and $U(k_1)$}&\qquad \hbox{ordinary}&\qquad
\hbox{$\Delta = m/g$}\cr
&\hbox{$U(N_1+k_1)$ and $U(N_2+k_2)$}&\qquad \hbox{ghost}&\qquad
\hbox{$\Delta = m/g$}\cr
&\hbox{$U(k_1)$ and $U(k_2)$}&\qquad \hbox{ghost}&\qquad \hbox{$\Delta
= m/g$}\cr
}}
The non-zero masses are seen by the same sort of analysis as in \dgkv.

We now ask where there can be residual instanton effects in Higgsing
between $k_1, k_2 = \infty$ and $k_1,k_2=0$.  Exactly as above, this
can happen when some $U(2)$ with $N_f=2$ massless flavors is Higgsed
to $U(1)$.  The $U(k_1)$ and $U(k_2)$ groups have too many massless
flavors, and the only possibility is when $N_1=k_1=1$ or $N_2=k_2=1$.
So we consider $U(N_2+k_2)$, with $N_2=k_2=1$; this gauge group is
$U(2)$ with $N_f=2k_2=2$ massless flavors, which leads to a
superpotential exactly as in \wsuiis.  The only difference from
\wsuiis\ and \wlowii\ is that the dynamical scale $\Lambda ^2$ there
should be regarded as the low energy scale, after we've integrated out
the massive bi-fundamentals in \bifundl.  Since $U(N_2+k_2)$, for
$N_2=k_2=1$, has $2k_1$ ordinary matter fields with mass $\Delta$ and
$2(N_1+k_1)$ ghost matter fields with mass $\Delta$, the correct
replacement is $m\Lambda ^2
\rightarrow m \Lambda ^{2N}\Delta ^{(2k_1-(2N_1+2k_1)}= m\Delta
^{-2N_1}\Lambda ^{2N}$.
The ordinary and ghost matter fields contribute with opposite sign in
the exponent because
of their opposite sign contributions to the one-loop beta function.

The upshot is that when we Higgs $U(N_1+N_2)\rightarrow U(N_1)\times
U(N_2)$ we get
a glueball superpotential
\eqn\wgbs{W=N_1S_1\log S_1+N_2S_2\log S_2+\dots,}
where the $N_i S_i \log S_i$ is present even for $N_i=1$; in that
case, it arises as described above, from a residual instanton term
when our large $k$ F-completion is Higgsed down to $k=0$.  For
example, if we consider $U(2)\rightarrow U(1)\times U(1)$, the
glueball superpotential contains terms $S_1\log S_1+S_2\log S_2$.

These residual instanton terms, which we just found for the particular
example of $U(N)\rightarrow U(N-1)\times U(1)$, are actually also present
in the standard gauge theory description of this breaking, for all
$N>1$.  We described the residual instanton contributions above as
arising from the F-completion of the $U(1)$ factor into $U(2|1)$.
But, in the standard gauge theory description of this breaking
pattern, these residual instanton contributions are also present,
coming from the broken part of the UV group, \ie\ $U(N)/U(N-1)\times U(1)$.
This is why the ``$U(1)$ instanton'' factor found above was
$m\Delta ^{-2N_1}\Lambda ^{2N}$, which indeed has the correct
$\Lambda$ exponent to be a $U(N)$ instanton.  As an example, in
$U(2)\rightarrow U(1)\times U(1)$, the residual instantons seen in our
 F-completion of the $U(1)$ factors are actually just the
ordinary instantons in the original $U(2)$, i.e. they are 
residual instantons associated with $\pi _3(U(2)/U(1)\times U(1))
\neq 0$.  The effect of these instantons in
$U(2)\rightarrow U(1)\times U(1)$ was, for example, already included
in the analysis of \CachazoPR, where the result was verified to agree
with the standard $U(2)$ gauge theory result of \SeibergRS.

\subsec{$Sp(N)$ with an adjoint and superpotential $W(\phi)$}

The $Sp(N+k|k)$ F-completion was discussed in sect. 4.3.  We first
consider the theory where the adjoint field is very massive and can be
integrated out, to describe the case of pure ${\cal N}=1$ $Sp(N)$
Yang-Mills.  As discussed above, we will see that the F-completion
includes a residual instanton contribution, which occurs only for
$N=0$ and is precisely that needed to
eliminate the non-smooth $\delta _{N,0}$ term in the standard UV
completion of gauge theory \hsgt.  The F-completion is smooth for all
$N$, and in particular leads to a non-zero superpotential,
given by \wsymb\ with $h=1$, for the F-completion of $Sp(0)$.

Integrating out the massive adjoints,  the F-completion gauge theory
is \spsgi, without the adjoints, and with a quartic tree-level
superpotential as in \wol.  We consider when there can
be residual instanton contributions in a Higgsing $k\rightarrow k-1$,
looking for non-trivial $\pi _3(\cG /\cH)$, with $\cG/\cH =
Sp(N+k)/Sp(N+k-1)$ or $SO(2k)/SO(2k-2)$.  The only non-trivial cases
are $Sp(1)/Sp(0)$, which occurs for $N=0$ and $k=1$, or $SO(4)/SO(2)$,
corresponding to $k=2$.  Since the $SO(2k)$ generally has $4(N+k)$
massless flavors, the $k=2$ case of $SO(4)/SO(2)$ always has far too
many massless flavors to lead to a residual instanton superpotential.
So the only possible case for a residual instanton contribution is
$N=0$, Higgsing $Sp(1)\times SO(2)$ to nothing.

So consider the $Sp(1)\times SO(2)$ theory.  We write the
bi-fundamentals as $(Q_i)_c^{c'}$, with $i=1,2$; $c$ is
the $Sp(1)$ fundamental index and $c'=1,2$ the $SO(2)$ fundamental
index.  The $Sp(1)\cong SU(2)$ has $N_f=2$, so the theory is described
by exactly the same superpotential as in \wsuiis.  We organize the
gauge invariants in terms of mesons $M^{c'd'}=(Q_1)_c^{c'}(Q_2)_d^{d'}
\epsilon ^{cd}$ and baryons $B=\det _{cc'}((Q_1)_c^{c'})$, $\widetilde
B=\det _{cc'}((Q_2)_c^{c'})$.  As described after \mbeom\ there are
two solution branches and, again, the vacuum with $\ev{M^{c'd'}}=0$ and
$\ev{B\widetilde B}=0$ is not the right one for our purposes, because it
not connected to the moduli space where the gauge group can be Higgsed
to $Sp(0)$ at an arbitrarily high scale.  As before, the correct
branch is $\ev{\det M}=mS$ and $\ev{B}=\ev{\widetilde B}=0$.
As in \wlowii, this leads to the usual gaugino condensation
glueball superpotential \wsymb, with coefficient $h=1$ for $Sp(N=0)$.
Again, this precisely corresponds to eliminating the non-smooth
behavior of \hsgt\ in favor of the smooth result $h=N+1$, for all
$N\geq 0$.

Now consider the $Sp(N+k|k)$  F-completion with an adjoint and
a quartic superpotential $W=Str W(\Phi)$, with $W(\Phi) = {g\over 4}
\Phi ^4
+{m\over 2}\Phi ^2$.  Now in the general vacuum we have
$Sp(N_1+N_2+k_1+k_2|k_1+k_2)\rightarrow Sp(N_1+k_1|k_1)\times
U(N_2+k_2|k_2)$, where $2N_1$ is the number of zero eigenvalues of
$\phi _1$ and $\phi _2$, and $2N_2$ is the number with eigenvalue
corresponding to the vacuum at $a = -m/g$.  Our F-completion of the IR
group is $Sp(N_1+k_1)\times SO(2k_1)\times U(N_2+k_2)\times U(k_2)$.
Taking into account the numbers of massless and massive bi-fundamental
flavors, as in \bifundl, we find that there are residual instanton
effects only in the cases of $N_1=0$, where the instanton comes when
we Higgs $k_1=1$ to $k_1=0$, and $N_2=1$ , where the instanton comes
when we Higgs $k_2=1$ to $k_2=0$.

So the residual instanton effects, associated with our F-completion,
are only present for $Sp(N)\rightarrow Sp(N-1)\times U(1)$ and for
$Sp(N)\rightarrow U(N)$.  For the case $Sp(N)\rightarrow Sp(N-1)\times
U(1)$, the residual instanton effects look like instantons in the
$U(1)$ factor, associated with the F-completion of $U(1)$.  But, as in
the $U(N)$ case, these same instanton effects are already present in
the standard gauge theory UV completion of $Sp(N)$; they are ordinary
instanton effects in the original $Sp(N)$.  There they again arise as
residual instanton effects, associated with the case $Sp(N)\rightarrow
Sp(N-1)\times U(1)$.  Likewise, for $Sp(N)\rightarrow U(N)$, one might
suspect there to be residual instanton effects associated with our
F-completion, because it's really $Sp(N)\rightarrow Sp(0)\times U(N)$,
and the F-completion of $Sp(0)$ to $Sp(1)\times SO(2)$ has a residual
instanton effect, as described above.  But, again, this same residual
instanton contribution is already present in the standard UV
completion of $Sp(N)$ gauge theory.  From that perspective it is a
residual instanton effect associated with $\pi _3(Sp(N)/U(N))=Z_2$.

So our F-completion of $Sp(N)$ with an adjoint differs from the
standard gauge theory UV completion for one, and only one, case:
$Sp(0)$.   We thus expect agreement
between the matrix model results, which corresponds to the large $k$
F-completion, and standard gauge theory results for $Sp(N)$ gauge theory
with an adjoint, for all Higgs breaking patterns, and for any $N>0$.

\subsec{$SO(N)$ with an adjoint and superpotential $W(\phi)$}

The F-completion to $SO(N+2k|k)$ was described in sect. 4.2.  We
consider first the case where the adjoint is massive and can be
integrated out.  So the gauge theory is the
ordinary $SO(N+2k) \times Sp(k)$ gauge
theory \sosgi, with the adjoints integrated out and the resulting
quartic superpotential \wol\ for the bi-fundamentals.  As before, we
can immediately determine when there can be an residual instanton
contribution to the superpotential in the Higgsing $k\rightarrow k-1$:
the only possibilities for a non-trivial $\pi _3(\cG/\cH)$ are $SO(4)
\rightarrow SO(2)$ and $Sp(1)$ breaking completely.  The $SO(N+2k)$ has
$N_f=4k$ vectors and can never lead to a superpotential.  So the only
possibility for a residual instanton effect is in the $Sp(k)$ factor,
when $k=1\rightarrow k=0$.  The $Sp(1)$ factor has $N_f=N+2$, and can
only lead to a superpotential when $N_f<4$, \ie\ $N=0$ and $N=1$.
Again, these are precisely the cases where the standard gauge theory
results are non-smooth, and we'll see that the F-completion has
precisely the correct residual instanton contributions needed to
smooth out \hsgt.

Consider first the F-completion of $SO(0)$ to $SO(2)\times Sp(1)$.
Higgsing the F-completion back down to $SO(0)$ leads to a residual
contribution, precisely as in the discussion of the F-completion of
$Sp(0)$ in the previous section.  There is, however, a small
difference from our previous examples: this time the instanton is not
in the $G(N+k)$ factor, namely $SO(2)$, 
but rather the other factor in $G(N+k|k)$, i.e. the instanton in Higgsing
$Sp(1)\rightarrow Sp(0)$.  As
discussed at the end of sect. 4.2, the $Sp(k)$ instanton factor is
actually the inverse square root of the $SO(N)$ instanton factor.  In
particular, the $Sp(1)$ instanton leads to a superpotential
contribution $W=-\Lambda ^3$, but the $SO(0)$ instanton factor is
naturally $\Lambda ^{3(0-2)}$.  We thus write the residual instanton
term in the standard glueball superpotential form \wsymb, with $h=-2$.
The minus sign comes essentially from that of the $Str$, since the
instanton is in the $Sp(k)$ factor of $SO(N+2k|k)$, and the $2$ is
because the $Sp(k)$ instanton is the inverse square root of the
$SO(N)$ instanton.  The result $h=-2$ is precisely the correct result
smoothing out $h(SO(N))=N-2$, to apply even for $N=0$.

The F-completion of $SO(1)$ involves a slight variant of the above
discussion, since now the F-completion is $SO(3)\times Sp(1)$, and
the $Sp(1)$ has $N_f=3$, rather than $N_f=2$, flavors.  Instead of
the quantum-modified moduli space constraint, this theory is
described by the superpotential \Seiberg
\eqn\suiinfiii{W=-{1\over m^2\Lambda }(M^{ab}B_{a}\widetilde B_{b}-
\det _{ab}M^{ab})+W_{tree},}
where $\Lambda$ is the scale of the theory including the adjoint of mass
$m$.   We organize the gauge invariants as $M^{ab}=(Q_1)_{a'}^{a}(Q_2)_{b'}^
{b}
\epsilon ^{a'b'}$ and $B_{a}=\epsilon _{abc}(Q_1)_{a'}^{b}(Q_1)_{b'}^{c}
\epsilon ^{a'b'}$, and $\widetilde B_{a}=\epsilon _{abc}(Q_2)_{a'}^{b}
(Q_2)_{b'}^{c} \epsilon ^{a'b'}$, with $a$ the $SO(3)$ index and $a'$
the $Sp(1)$ fundamental index.  We can write the tree-level quartic
superpotential, obtained by integrating out the massive adjoints, as
\eqn\wtreexi{W_{tree}=-{1\over m}(B_{a}\widetilde B^{a}-
M_{ab}M^{ab}).}

Integrating out the fields $M$ and $B$ we have a moduli space of
vacua, along which we can Higgs to $SO(1)$, with the constant
low-energy superpotential
\eqn\wlowxi{W=m\Lambda ^2=\Lambda _L^3,}
where $\Lambda _L$ is the scale of the low energy $SO(N)$ Yang-Mills
theory.  In terms of the $Sp(1)$ theory, the superpotential \wlowxi\
has the quantum numbers of a two-instanton contribution.  But, because
the $Sp(k)$ is the inverse square-root of the $SO(N)$ instanton
factor, the superpotential \wlowxi\ looks like the inverse of an
$SO(N)$ instanton.  In particular, the $SO(N)$ instanton factor is
$\Lambda ^{3(N-2)}=\Lambda ^{-3}$ for $N=1$.  So the residual instanton
term \wlowxi\ can be described via the usual glueball superpotential \wsymb,
but with $h=-1$.  This is precisely the value associated with the
smooth result \wlowuiv, continued to $N=1$.

Continuing on to $SO(N)$ with an adjoint and more general superpotential,
the discussion is very similar to that of the $Sp(N)$ theory discussed
in the previous subsection.  For example, with a superpotential including
a quartic term, we break $SO(N+2M)\rightarrow SO(N)\times U(M)$ in the
general vacuum.  Our F-completion of this is $SO(N+2M+2k_1+2k_2|k_1+k_2)
\rightarrow SO(N+2k_1|k_1)\times U(M+k_2|k_2)$.  The residual instantons
only occur for $k_1=1$, with $N=0$ or $N=1$ as in the above discussion,
or for $M=1$, with $k_2=1$, \ie\ $SO(2M)\rightarrow SO(0)\times U(M)$, $SO(2M+1)
\rightarrow SO(1)\times U(M)$, and
$SO(N+2)\rightarrow SO(N)\times U(1)$.  In all of
these cases, these residual instanton contributions from our
F-completion are already present in the standard UV completion of
$SO(N+2M)\rightarrow SO(N)\times U(M)$. For example, the standard UV
completion gives a residual instanton associated with $\pi _3(G/H)$
when we break $\cG =SO(2M)$ to $\cH =U(M)$.

\subsec{$Sp(N)$ with an antisymmetric tensor $\phi$ and superpotential
$W(\phi)$}

The F-completion was discussed in sect. 4.5.  When the superpotential
is a mass term for the field $\phi$, the low energy theory is the same as
with $Sp(N)$ with a massive adjoint, so we already know what happens
in that case: there is a residual instanton term only for $N=0$, making
$Sp(0)$ contribute as expected based on smoothly applying $h=N+1$ for
all $Sp(N)$, including $N=0$.

On the other hand, for higher order in $\phi$ tree-level superpotentials,
$W(\phi)= \sum _{p=1}^{n+1}{1\over p}g_p 
\Tr (\phi J)^p$, with $n>1$, our F-completion
contains additional residual instanton contributions, which are not
already present in the standard UV completion of these gauge theories.
This sets these examples apart from those of the previous sub-sections.
This could, in principle, lead to a discrepancy with standard gauge theory,
as was found in \KrausJF.

To motivate the additional residual instanton
contributions,
consider the simplest interesting case, a cubic superpotential:
$W=\Tr ( {g\over 3}(\phi J)^3+{m\over
2}(\phi J)^2)$.  In the general vacuum, we break $Sp(N_1+N_2)\rightarrow
Sp(N_1)\times Sp(N_2)$.  Our F completion of this is
$Sp(N_1+N_2+k_1+k_2|k_1+k_2)
\rightarrow Sp(N_1+k_1|k_1)\times Sp(N_2+k_2|k_2)$.  The IR theory can be
gauge fixed to a quiver
gauge theory with gauge group $Sp(N_1+k_1)\times
SO(2k_1)\times Sp(N_2+k_2)\times SO(2k_2)$, with two bi-fundamentals
connecting every pair of gauge groups, with mass and statistics
given by
\eqn\spbifundl{\matrix{&\hbox{\bf Groups connected}&\qquad \hbox{\bf
statistics}&\qquad\hbox{\bf mass}\cr
&\hbox{$Sp(N_1+k_1)$ and $SO(2k_1)$}&\qquad \hbox{ordinary}&\qquad
\hbox{massless}\cr
&\hbox{$Sp(N_2+k_2)$ and $SO(2k_2)$}&\qquad \hbox{ordinary}&\qquad
\hbox{massless}\cr
&\hbox{$Sp(N_1+k_1)$ and $SO(2k_2)$}&\qquad \hbox{ordinary}&\qquad
\hbox{$\Delta = m/g$}\cr
&\hbox{$Sp(N_2+k_2)$ and $SO(2k_1)$}&\qquad \hbox{ordinary}&\qquad
\hbox{$\Delta = m/g$}\cr
&\hbox{$Sp(N_1+k_1)$ and $Sp(N_2+k_2)$} &\qquad \hbox{ghost}&\qquad
\hbox{$\Delta = m/g$}\cr
&\hbox{$SO(2k_1)$ and $SO(2k_2)$}&\qquad \hbox{ghost}&\qquad \hbox{$\Delta
= m/g$.}\cr}}
The instanton factors for $Sp(N_1+k_1)$ and $Sp(N_2+k_2)$ are
\eqn\spspmr{\Lambda _1^{3(N_1+1)}=g^{N_1-1}\Lambda ^{2N+4}\Delta ^{N_1-1-2N_2},
\qquad \Lambda _2^{3(N_2+1)}=g^{N_2-1}\Lambda ^{2N+4}(-\Delta)^{N_2-1-2N_1}.}

Taking $N_1=N$ and $N_2=0$, we should expect additional, residual instanton
contributions to $W$, coming from the $Sp(k_2|k_2)$ completion of the $Sp(0)$ factor.
Since the residual instanton contributions occur in the last step of
Higgsing $k\rightarrow k-1$, it suffices to consider the F-completion to
just $Sp(N+1|1)\rightarrow Sp(N)\times Sp(1|1)$, \ie\
$Sp(N+1)\times SO(2)\rightarrow  Sp(N)\times Sp(1)\times SO(2)$.
The matter content is as in \spagi, with the interesting
dynamics, which leads to the residual instanton contributions to
the superpotential, that associated with the $Sp(N+1)$ factor.
The $Sp(N+1)$ theory in \spagi\ has an antisymmetric tensor $\phi$
and $N_f=2$ flavors (four fundamentals).  That theory has a quantum
moduli space, with one of the classical constraints modified by a
$Sp(N+1)$ instanton, and $N$ classical constraints
unmodified; see \refs{\ChoBI , \CsakiEU}
for a general discussion, and explicit analysis of some low $N$
examples.  To this we can add the tree level superpotential in \wspuu\
and explicitly compute the full low -energy superpotential, including
all residual instanton corrections to the original $Sp(N)$ theory.

As an example, consider the simplest case, where the original theory
is $Sp(1)$ with an anti-symmetric.  Since the anti-symmetric is a
gauge singlet, this is just $Sp(1)\cong SU(2)$ super-Yang-Mills, with
an additional decoupled singlet. This case was not explicitly
discussed in \KrausJF, because the gauge theory appears to be too
trivial, with the $SU(2)$ dynamics completely decoupled from the
parameters in $W_{tree}$.  But the matrix model results of \KrausJF\
apply to this case as well, we just set $N=2$ in their matrix model
formulae, which yields non-trivial results.  We interpret these as
applying to the $Sp(1+k|k)$ F-completion \spagi, whose residual
instanton contributions are already apparent for the case $k=1$,
\ie\ $Sp(2)\times SO(2)$.  The relavant $Sp(2)$ gauge theory has
gauge invariants ${\cal O}_1\equiv \half \Tr (\phi J)$,
${\cal O}_2=\half \Tr (\phi J)^2-{\cal O}_1^2$, 
$M_{[ij]}=Q_iQ_j$, and $N_{[ij]}=
Q_i \phi Q_j$, with $i,j=1\dots 4$.  The superpotential is (we modify
the results of \refs{\ChoBI , \CsakiEU} to include the trace ${\cal O}_1$.)
\eqn\wmess{W=\lambda _1(2N_{ij}N^{ij}+
M_{ij}M ^{ij} {\cal O}_2-2\Lambda ^6)+\lambda _2(M_{ij}N^{ij}
- \half {\cal O}_1 M_{ij}M^{ij}) +W_{tree},} 
where $W_{tree}$ are all of
the terms in \wspuu.  Here $\lambda _1$ and $\lambda _2$ are Lagrange
multipliers and we use $\epsilon ^{ijkl}$ to raise indices, so $N_{ij}N^{ij}
\equiv Pf (N)$.  Writing $W'=gx(x+m/g)$, the classical
eigenvalues of $\phi$ and the $SO(2)$ symmetric tensor $\phi '$ 
are $0$ and $-m/g$.  We are interested in
the vacuum where both $\phi $ and $\phi '$ classically have one
eigenvalue zero and one $-m/g$, breaking $Sp(2|1)\rightarrow
Sp(1)\times Sp(1|1)$.  We leave a full analysis of the vacuum and
residual superpotential to future work.  This is currently under investigation.

\subsec{$U(N)$ ${\cal N}=1^*$}

The $U(N+k|k)$ F-completion has
the exact effective superpotential
\eqn\wstarm{W_{mm}=-{Nm^3\over 12}E_2(\tau
)=Nm^3(q+3q^2+4q^3+7q^4+\dots ),}
where $\tau= \tau _0/N$ and $q=e^{2\pi i \tau}$, with the instanton
factor $q^N$.
On the other hand, the standard gauge theory result is
\eqn\wstarg{W_{gt}=-{Nm^3\over 12}E_2(\tau )+{N^2m^3\over
12}E_2(N\tau).}  The added term contributes at instanton order and
higher, so this discrepancy could perhaps be related to residual
instanton terms, associated with our F-completion.

In particular, consider  $U(1)$ ${\cal N}=1^*$, where the
standard gauge theory result \wstarg\ gives zero non-perturbative
superpotential, as might have been expected for this free theory.  On
the other hand, the result \wstarm\ is non-zero.  We would like to
interpret this difference as coming from residual instanton
contributions in breaking $U(1+k|k)$ to $U(1)$; in particular, it
comes from the Higgsing of $U(2|1)$ to $U(1)$.  Clearly, there will
indeed be such residual instanton contributions in this case: we can
integrate out the massive fields in our F-completion \unssgi\ of the
${\cal N}=1^*$ theory to obtain a low energy theory which coincides
with our F-completion of the ${\cal N}=1$ $U(1)$ theory.  And we
already saw that our F-completion of the ${\cal N}=1$ $U(1)$ theory has
a residual instanton contribution $\Lambda ^3$, with coefficient 1.
Here $\Lambda$ is the scale of our ${\cal N}=1^*$
F-completion \unssgi\ with the massive fields integrated out, so
$\Lambda ^3=m^3q$ in this $U(1)$ case.  We thus precisely recover
the $q$ term in \wstarm, with the correct coefficient 1.  It should 
be possible to also recover the higher instanton terms in \wstarm;
since the ordinary gauge theory result for $U(1)$ is $W=0$, 
the full set of residual instantons should sum to $m^3E_2(\tau)/12$.
Explicitly computing these, and the higher $N$ generalizations,
{}from our F-completion is more challenging.  The challenge is in how to 
properly include the effects of the additional massive matter fields
in \unssgi, because some are fermionic; this is currently
under investigation.

\newsec{Perturbative expansion for $\cN=1^*$  models}

Here we apply the formalism of section 3 to ${\cal N}=1^*$ theory
in the maximally confining phase. For simplicity we will focus on the
simply-laced Lie algebras\foot{Apart from having all the
roots with (length)$^2 = 2$, one should also remember that the
Coxeter number and dual Coxeter number are equal for such Lie algebras.}.
For the $\cN=1^*$ models the weights,
$\vec \lambda_a$ are in fact the roots, $\vec \alpha$, of the Lie
algebra.  We will focus on the group dependence of the Feynman
diagrams and use the known answer for the $SU(N)$ case for fixing the
combinatoric prefactors of the diagrams.

At one loop the determinant in \graphcont\ is simply the
length-squared of a root, that is, $2$.  The number of roots in a Lie
algebra is $h \, r$, where $h$ is the Coxeter number and $r$ is
the rank.  Thus, summing the determinant over the roots yields $2 h\,
r$.  The factorial $(r -\ell)!/r!$ gives a factor of $1/r$, for an
overall contribution of $2h$.

The two-loop diagram is shown in Figure 1.     Let $\vec \alpha$ and
$\vec \beta$ denote the root vectors running around each loop.  The
group
theory structure constants are then given by the structure constants in
the root basis:
$$
\big[ E_\alpha \,, \, E_\beta \big]  ~=~ {N_{\alpha \, \beta}}^\gamma \,
   E_\gamma \,.
$$
In a simply-laced Lie Algebra, $ {N_{\alpha \, \beta}}^\gamma$ is
non-zero if and only if $\vec \alpha\cdot \vec \beta = -1$, and in
standard normalization the $E_{\alpha}$'s can be chosen so that the $
{N_{\alpha \, \beta}}^\gamma$'s are $\pm 1$.  Since there are two
vertices in the two-loop graph, it follows that the group theory
factors is $ ({N_{\alpha \, \beta}}^\gamma)^2$ which is either zero or
one depending on whether $\vec \alpha\cdot \vec \beta = -1$.  Thus we
only have to sum the determinant:
$$
\det \, \left( \matrix{ \vec \alpha\cdot \vec \alpha &  \vec \alpha\cdot
   \vec \beta  \cr
   \vec \beta \cdot \vec \alpha  &  \vec \beta \cdot \vec \beta } \right)
~=~
   \det \, \left( \matrix{2 &-1 \cr -1& 2} \right) ~=~ 3
$$
over all the roots with $ \vec \alpha\cdot \vec \beta = -1$.

\goodbreak\midinsert
\vskip .2cm
\centerline{ {\epsfxsize 1.5in\epsfbox{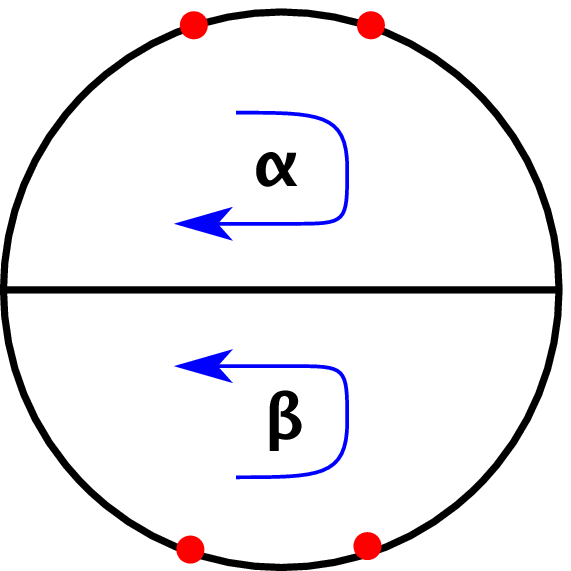}}}
\vskip .25cm
\leftskip 2pc
\rightskip 2pc\noindent{\ninepoint\sl \baselineskip=8pt
{\bf Fig.~1}:
The two loop diagram showing the gauge theory charges,
$\vec \alpha, \vec \beta$ in each loop.  The dots denote
the insertions of ${\cal W}_\alpha$.
}
\endinsert

To evaluate this sum, recall that in the root basis the quadratic
Casimir operator
on adjoint is:
$$
\cC_2 (X)~=~ \sum_{I=1}^r \, \big[\, H_i \,, \big[ \,H_i \,,  X \,\big]
\,\big]  ~+~
{1 \over 2}\,\sum_{ \alpha} \, \big[  \,E_{-\alpha}  \,, \big[\,
E_\alpha  \,,  X\, \big]   \,\big]
~=~ h \, X   \,.
$$

Evaluating this on $E_\beta$, using the properties of the ${N_{\alpha
\, \beta}}^\gamma$
gives:
$$
\vec \beta \cdot  \vec \beta ~+~ {1 \over 2} \, \sum_{\alpha: \vec
\alpha \cdot
\vec \beta = -1} \,  1  ~=~ h\,,
$$
and hence, for a given root, $\vec \alpha$, there are precisely
$2(h-2)$ roots
that have inner product $-1$ with it.  Since the total number of roots
is $h r$, the group theory factor for the two loop diagram is $ 3
\times 2 h (h-2) r$.
Thus \graphcont\ at two loops gives:
\eqn\twoloops{\big( \Delta W \big)_{\cG_2} ~=~ {6\, h \, (h-2) \over
(r-1) } \, S^2 \,. }

There are two diagrams at three loops, a ``double-bar'' graph, and a
``peace-sign''
graph.  These are shown in figure 2.  The  ``double-bar'' graph involves
three roots $\vec \alpha$, $\vec \beta$ and $\vec \gamma$ that satisfy
$\vec \alpha \cdot   \vec \beta = -1$ and $\vec \beta \cdot   \vec
\gamma = -1$.
There is no constraint on $\vec \alpha \cdot   \vec \gamma$  except
that linear
independence requires that this inner product not be $\pm 2$ of $-1$ (if
it is $-1$ then $\vec \alpha + \vec \beta+ \vec \gamma =0$).  Thus
$\vec \alpha \cdot   \vec \gamma$ can be $0$ or $+1$.
We therefore have two types of root triples:

\item{(i)}  $(\vec \alpha, \vec \beta ,\vec \gamma)$   with
$\vec \alpha \cdot   \vec \beta = -1$,  $\vec \beta \cdot   \vec \gamma
= -1$
and $\vec \alpha \cdot   \vec \gamma =0 $
\item{(ii)}  $(\vec \alpha, \vec \beta ,\vec \gamma)$   with
$\vec \alpha \cdot   \vec \beta = -1$,  $\vec \beta \cdot   \vec \gamma
= -1$
and $\vec \alpha \cdot   \vec \gamma =+1$.

Suppose that $(\vec \alpha, \vec \beta ,\vec \gamma)$ is a type (ii)
triple, then define $\vec \gamma' = -( \vec \beta + \vec \gamma)$,
then $(\vec \alpha, \vec \beta ,\vec \gamma')$ is a type (i) triple.
It thus follows that the numbers of each type of triple within a Lie
algebra are equal.  Also note that the determinant in \graphcont\ is
equal to $4$ for each type of triple.  The ``peace-sign'' graph
requires that $(\vec \alpha, \vec \beta ,\vec \gamma)$ be a triple of
type (ii).

We therefore need to count all triples of type (i) in our Lie algebra,
$\cL$.  We computed these explicitly, however we subsequently found a
simpler way to encode the result.  Observe that a triple of type (i)
defines an $A_3$ (or $SU(4)$) subalgebra of $\cL$, and so the problem
amounts to counting all regularly embedded $A_3$ subalgebras.  Our
explicit computation shows that one can do this from the extended
Dynkin diagram as follows: Take $\vec
\alpha$ to be
the extending node and then  take $\vec \beta$ and $\vec \gamma$ to be
connected nodes on the extended Dynkin diagram.
Delete these nodes from the diagram, {\it and}   delete all nodes that
are connected
to the $\vec \alpha$, $\vec \beta$ and $\vec \gamma$ nodes.  Call the
residual
Lie Algebra, $K$.    It turns out the number of different $A_3$
subalgebras is
then $|W_\cL |/|W_\cK|$, where $W_\cX$ is the Weyl group of $\cX$ and
$|W_\cX|$ is its order.  If there is more than one way to
choose the $\vec \alpha$, $\vec \beta$ and $\vec \gamma$ nodes
(starting with $\alpha$ as the extending root) then one sums over
the corresponding $|W_\cL |/|W_\cK|$.

\goodbreak\midinsert
\vskip .5cm
\centerline{ {\epsfxsize 4in\epsfbox{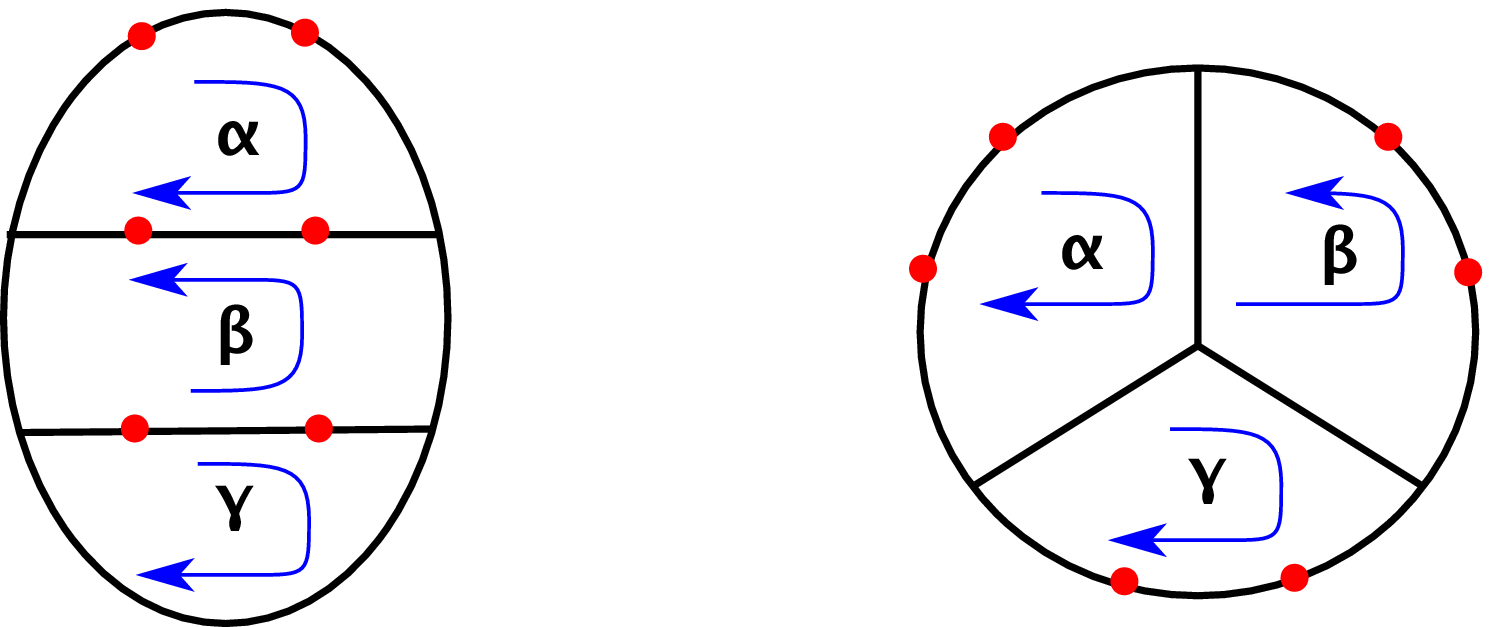}}}
\leftskip 2pc
\rightskip 2pc\noindent{\ninepoint\sl \baselineskip=8pt
{\bf Fig.~2}: The two possible three loop graphs showing
the gauge theory charges $\vec \alpha,\vec \beta$ and $\vec \gamma$.
}
\endinsert

We then arrive at the following numbers of triples of type (i):
\eqn\triples{\eqalign{ A_{N-1}: & \quad 2\, N\,(N-1)\, (N-2) \, (N-3)
\,,  \cr
D_{N}:  &  \quad 8\, N\,(N-1)\, (N-2) \, (2N-5) \,,  \cr
E_{6}:    &\quad 2 \,. 3^2  \,. 6! \,,   \qquad \quad
E_{7}:   \quad 2^2 \,.3 \,. 7! \,,   \qquad \quad
E_{8}   \quad  3^2 \,. 8!      \,.}}
These have to be multiplied by $4$ because of the determinant, and by
the factor of ${(r-\ell)! \over r!}$ in \graphcont.  There is also a
factor of
$2$ for the ``double-bar'' graph because  triples of both types
contribute.
Finally there are symmetry factors for the graphs and the ${1 \over 4!}$
{}from the Feynman diagram expansion.  This gives a ${3 \over 4}$ to the
``double-bar'' graph and a ${1 \over 4}$ to the ``peace-sign.''  The
foregoing group factors therefore get muliplied by $ 4 \times (2 \times
{3 \over 4}
+ {1 \over 4}) {1 \over r (r-l)(r-2)} =  {7 \over r (r-l)(r-2)}$.  The
end  result is then a contribution to  $W(s)$ of:
\eqn\threeloopgp{\eqalign{ A_{N-1}: & \quad 14\, N \, S^3 \,,  \qquad
\quad
D_{N}:    \quad 14\times 4 \, (2N-5) \, S^3\,,    \cr
E_{6}:    &\quad 14\times 54\,  S^3 \,,   \qquad \quad
E_{7}:   \quad 14 \times 144 \, S^3\,,    \qquad \quad
E_{8}   \quad  14\times 540 \, S^3    \,.}}

While we have faithfully reproduced the result of \dgkv\ for $SU(N)$,
it is
important to realize that {\it graph by graph}, the results for other
groups
are obtained from the $SU(N)$ result by replacing $N$ by integers that
depend on the group and the loop order.  These integers are shown in
Table 1.

   \goodbreak
\bigskip
{\vbox{\ninepoint{
$$
\vbox{\offinterlineskip\tabskip=0pt
\halign{\strut\vrule#
&~$#$~\hfil\vrule
&~$#$~\hfil\vrule
&~$#$~\hfil\vrule
&~$#$\hfil
&\vrule#
\cr
\noalign{\hrule}
&
{\rm Gauge\  Group,} \ G
&
   \ell=1
&
\ell=2
&
\ell=3
&\cr
\noalign{\hrule}
&
A_N
&
N
&
N
&
N
&\cr
&
D_N
&
2N-2
&
4(N-2)
&
4(2N-5)
&\cr
&
E_6
&
12
&
24
&
54
&\cr
&
E_7
&
18
&
48
&144
&\cr
&
E_8
&
30
&
120
&
540
&\cr
\noalign{\hrule}}
\hrule}$$
\vskip-10pt
\noindent{\bf Table 1:}
{\sl   To obtain the contribution of an $\ell$-loop graph to $W(S)$ with
gauge group, $G$, the factors of $N$ for the $SU(N)$ result are to be
replaced by the
integers from this table.}
\vskip10pt}}}

These integers are, in fact, well known for each gauge group.
Consider the extended Cartan matrix, $\widehat C_{ij}$, where $i,j =
0, \dots, r$, where $0$ refers to the extending node.  This matrix has
a null vector, $\vec p
\equiv (p_0, p_1, \dots, p_r)$, whose entries are integers, and are
shown in Figure 3.  The integers appearing in Table 1 are then
precisely the integers:
\eqn\ppowers{ N_\ell ~=~ \sum_{j=0}^r \, (p_j)^\ell \,.}
Therefore, we find that the $\ell$-loop result (for $\ell \le 3$) is
given,
{\it graph by graph} by replacing:
\eqn\Srepl{ N S^\ell ~\rightarrow~ {\rm Tr}\big( \cS^ \ell) \,,}
where $\cS$ is a diagonal matrix whose entries are:
\eqn\Srepldef { \cS ~\equiv~ S\, {\rm diag}\big(p_0,p_1,\dots, p_r
\big)\,.}
We make the obvious conjecture that this is true to all loops for
simply-laced
groups.  It is also natural to ask what this F-completion prescription means
in terms of gauge theory.  At least for the $D$ series we can compare
this notion of F-completion with that given before based on supergroups.
We expect the amplitude to have a linear term in $N$ and a constant
piece, related to ${\bf CP}^1$ and ${\bf RP}^2$ contributions respectively.
The above conjecture for all loop answer agrees with this structure.
It would be interesting to check this explicitly.

\goodbreak\midinsert
\vskip .5cm
\centerline{ {\epsfxsize 2.5in\epsfbox{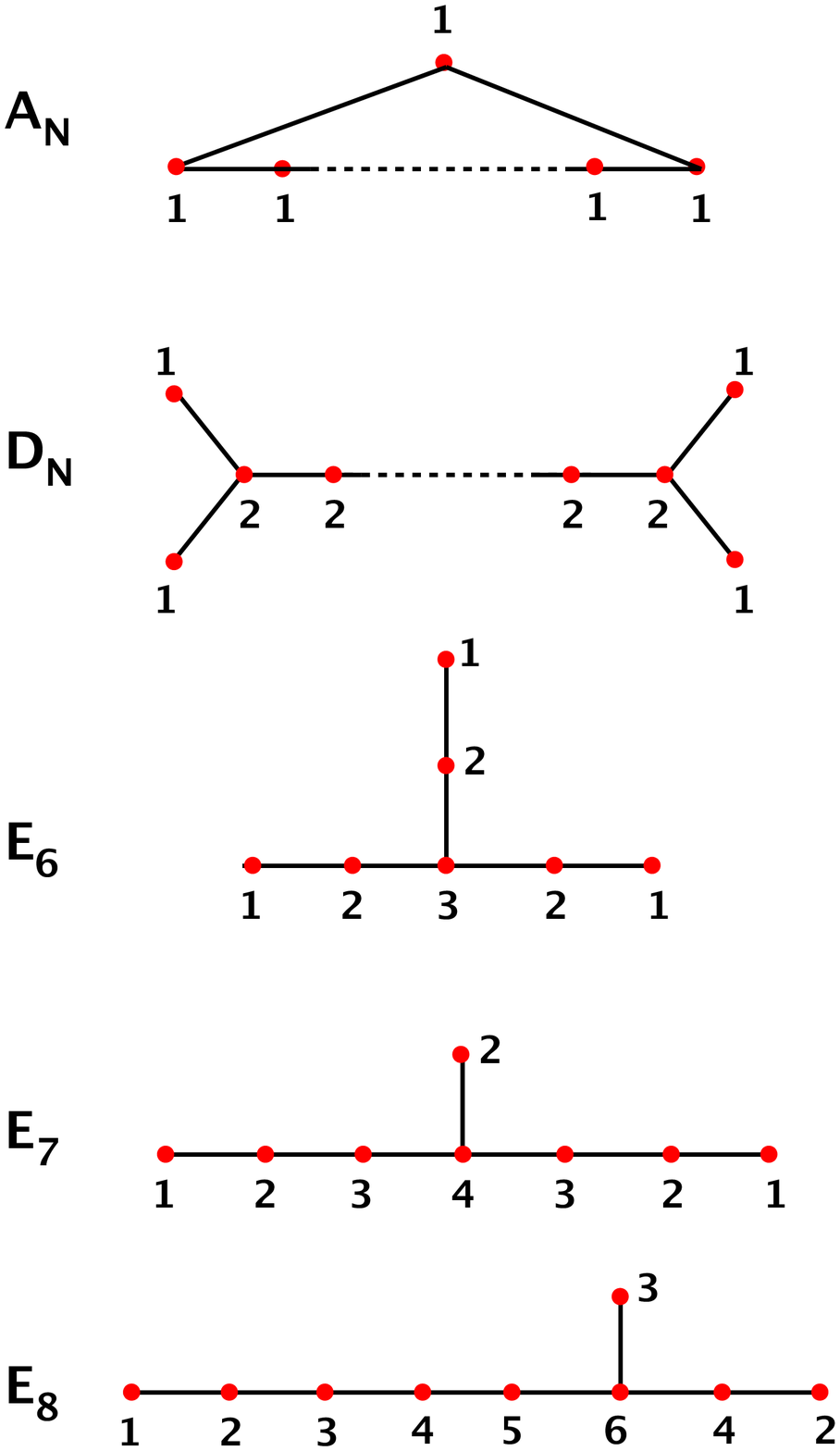}}}
\leftskip 2pc
\rightskip 2pc\noindent{\ninepoint\sl \baselineskip=8pt
{\bf Fig.~3}:
The extended Dynkin diagrams of the simply-laced lie algebras.
The integers, $p_a$, on each node are the Dynkin indices that
 define the null vector
of the extended Cartan matrix.
}
\endinsert

\newsec{Torus compactifications}

\subsec{ Compactification on $T^2$: A two-dimensional perspective}

There is an interesting link between ${\cal N}=1$ supersymmetric gauge
theories in four dimensions and certain ${\cal N}=2$ sigma models in
two dimensions.  The idea, which has been noted before \refs{\bjsv,
\HarveyTG, \hv, \LosevTU}, arises as follows: Consider
compactification of the pure ${\cal N}=1$ supersymmetric Yang-Mills
theory on $T^2$.  We can now turn on a Wilson line on $T^2$. The
F-term data do not depend on the volume of $T^2$.  Thus we can relate
the small volume description of F-terms, for which there would be an
effective ${\cal N}=2$ theory in two dimensions, to a large volume
description, which is effectively a four-dimensional, ${\cal N}=1$
supersymmetric F-term computation.  In the small volume limit of the
$T^2$, we have a good description of the two-dimensional theory as a
supersymmetric sigma model on the moduli space of flat connection of
the corresponding group on $T^2$.  Moreover the K\"ahler class of the
sigma model is identified with the coupling constant of the
four-dimensional theory.

The moduli space of flat connection on $T^2$ has been obtained in
\loijenga, and is given by the weighted projective space with
weights given by the Dynkin numbers $(p_0,...,p_r)$ of the
corresponding affine Dynkin diagram, where $p_0=1$ corresponds to
the extending root.  Note that this is a space of complex dimension $r$
as is expected.  In this map, the chiral ring of the gauge theory gets
identified with the chiral ring of the sigma model ({\it i.e.}  the
observables of the A-model topological strings).  The statement that
the K\"ahler class of the sigma model gets identified with the
coupling constant means that the glueball chiral field $S$, which
multiplies the bare gauge coupling in the action, gets identified with
the chiral field corresponding to the K\"ahler class of the sigma
model.

The sigma model can be realized in terms of a linear description
inolving a $U(1)$ gauge theory with $r+1$ chiral fields with charges
$(p_0,...,p_r)$.  Let $\Sigma$ denote the two-dimensional gauge field
strength multiplet, corresponding to the K\"ahler class.  In going
{}from four dimensions to two dimensions, the glueball superfield $S$
gets mapped to $\Sigma$; there is a bare term $\int \tau \Sigma\, d^2
\theta$ superpotential term.  Integrating out the matter fields leads
to a one loop induced superpotential for $\Sigma$:
\eqn\supe{W(\Sigma) ~=~ \Big(\sum_a p_a \Big) \, \Sigma \,  {\rm log}
( \Sigma - 1) ~-~ \tau \Sigma\,}
This agrees with the VY superpotential where we use the fact
that
$$\sum_{a=0}^r  p_a=h$$
is the dual Coxeter number.

Thus here we also find that $\Sigma^h=e^{-\tau}$ holds as a ring
relation.  As $\tau \rightarrow \infty$ this leads to $\Sigma^h=0$ as
the classical ring relation.  This relation may appear surprising at
first sight, because the dimension of the manifold is $r$ and one
would have naively expected that $\Sigma^{r+1}=0$ to be the ring
relation.  However, this is not quite true because the weighted
projective space is not smooth.  In particular there is a contribution
to $\Sigma$ from the twist fields at the singular loci, and they
violate the naive expectation that $\Sigma^{r+1}=0$. Similarly, it also
appears that the Witten index of this supersymmetric theory is $r+1$.
However this is not quite correct: The weighted projective space with
unequal weights has singularities.  There are contributions to the
index from the twisted sector and these contributions raise the naive
result, $r+1$, to the correct value, $h$.

For example, the weighted  projective space  for $SO(2n)$
has weights $1$ and $2$ with multiplicity $4$ and   $(n-3)$
respectively.
There is a twisted $Z_2$ sector, where the fixed point locus is
a $CP^{n-4}$.   In addition to the $n+1$ cohomology classes of the
underlying
space (\ie\ from the untwisted sector) we also get $n-3$ classes
{}from the twisted sector, from the cohomology of $CP^{n-4}$.  Altogether
this gives $2n-2$ cohomology classes which is the dual Coxeter
number of $SO(2n)$.  This generalizes to arbitrary group.  Consider
a simple group.  Let $n_k$ denote the number of Dynkin indices
of the affine Dynkin diagram with Dynkin index $k$.  Suppose that
all $k\not =1$ are relatively prime.
Then there is a $Z_k$ twisted sector which contributes $(k-1)$ twisted
sectors,
each with the cohomology of $CP^{n_k-1}$, \ie\ $n_k$ states. The
untwisted
sector contributes $r+1= \sum_k n_k$ to the cohomology.
    Thus the totality of contributions to the cohomology is given by
\eqn\sumup{\sum_k n_k+(k-1)\, n_k=\sum_k k \, n_k ~=~ h}
as expected.  When some of the $k$'s are not relatively prime the
counting still works the same way but the reasoning is slightly
different.  For example, suppose we have $n_2$ nodes with label 2
and $n_4$ nodes with label 4. Then there are three twisted sectors
related to $Z_4$, given by $4^{\rm th}$ roots of unity. However
the one given by the second root of unity is also part of the
$Z_2$ sector.  This implies that the two primitive fourth roots
give rise to a $CP^{n_4-1}$ fixed space, but the twisted sector
associated to $-1$ gives a $CP^{n_2+n_4-1}$ fixed locus.
Altogether this gives $2 n_4+ (n_2+n_4-1)=3n_4+n_2$ contribution
to the cohomology, exactly as in \sumup . The extra contributions
to the Witten index in this case are related to the puzzle raised
in \wiind\ that the Witten index of pure $N=1$ Yang-Mills is
naively equal to $r+1$.  This was resolved in the second paper in
\wiind\ where
specific flat connections on $T^3$ were identified as the source
of the discrepancy.  Note that $T^3=T^2\times S^1$ and if we take
$T^2$ to be small, we are discussing the sectors of the above
sigma model on $S^1$.   The twisted sectors we have found are related to
these extra contributions to the index.

It is also useful to consider using the mirror of this weighted
projective
space \refs{\hv,\av}.
We dualize the charged fields into chiral fields, $Y_a$,  from which we
obtain the superpotential:
$$W=\Sigma \, \Big(\sum_a p_a Y_a ~-~ \tau \Big)  ~+~ \sum e^{-Y_a}
\,.$$
Integrating out the $Y_a$ gives the superpotential \supe .
If we integrated out $\Sigma$ we get $W$ in terms of the affine
Toda superpotential, that is:
$$W=\sum_a e^{-Y_a} \,,  \qquad {\rm where} \quad \sum_a p_aY_a=\tau
\,.$$
That is, we get
$$W=\sum_{a\not=0} e^{-Y_a}+e^{-\tau} \prod_{a\not=1} e^{p_aY_a}\,. $$

This is also related to a three-dimensional description of F-terms,
as was noted in \kv\ and generalized to many other examples in  \dohol.
Namely we can also consider compactification from 4 down to 3 on a
circle.   In this case the $Y_a$ have the interpretation of the complex
field composed of gauge holonomies around the $S^1$ combined with
the dual of the gauge field in three dimensions, which is a scalar.

So far we have discussed the reduction of pure Yang-Mills.  If
we have some massive charged fields we can integrate them out
and get some correction to the foregoing superpotential.  It is natural
to expect that this can be done in two steps:  First integrating
out the charged fields, whose mass will vary with the choice of
flat connection, and then dualizing the flat conections to chiral fields
$Y_a$ as above. This would give a theory involving the neutral fields
interacting with the $Y_a$'s
defined above.
It would be interesting to work this out \foot{As we were
in the process of completing this work, an interesting paper appeared
\BoelsFH\ which gives the proposal of how this superpotential should be
modified for a single adjoint $U(N)$ theory, which we interpret
as the effect on the superpotential after
integrating out the massive charged fields.  Their result
suggests that in this case the massive charged modes are replaced by a
specific quantum vev.  See also the very recent work 
\ref\MAAM{Mohsen Alishahiha, Amir E. Mosaffa, ``On Effective 
Superpotentials and Compactification to Three Dimensions,
hep-th/0304247.}}.  
Integrating out all fields but the $S$ we will get an effective
superpotential as a function of $S$, $W(S)$. If we use our dictionary
this means that this computes a deformation of the two-dimensional
superpotential by the same function, that is, we get $W(\Sigma)$ in
the two-dimensional theory. Note that here the monomials in the
reduced superpotential $W_R(\Sigma)$ (\ie\ $W(\Sigma)$ up to the
$\Sigma^h$ term) get mapped to observables of the chiral ring in the
two-dimensional sigma model.  The quantum ambiguity $W_A(\Sigma)$ can
in principle be absorbed to a redefinition of the couplings.  It is
amusing to note that $W_A(\Sigma)$ would naturally correspond to
``gravitational descendants'' of the reduced phase space \losev\ where
the notion of ``gravitational descendent'' is defined in the context
of two-dimensional topological gravity \wittog.

\subsec{Compactification to three dimensions and ${\cal N}=1^*$
theories}

For the ${\cal N}=1^*$ theories with a simply-laced gauge group $G$,
the exact superpotential was conjectured in \Doreya , by
means very different from ours.

In \refs{\DoreySJ ,\dohol}\ the theory is considered on ${\bf
R^3}\times {\bf S^1}$,
where the light degrees of freedom of the theory can be written in
terms of $r = {\rm rank}(G)$ chiral superfields $Y_{1,\ldots, r}$ which live
on a torus of complex structure $\tau = {8 \pi^2\over {g_{YM}^2}} -i
\theta$
\SW.   It is argued that,
in these variables, the superpotential is an elliptic function on
the torus, namely
\eqn\ds{W(\vec Y; \tau) = m^3 \sum_{\vec{\alpha}> 0} {\cal
P}(\vec{\alpha}
\cdot \vec{Y} ),}
where $m$ is the mass used to deform the ${\cal N}=4$ theory to ${\cal
N}=1^*$, ${\cal P}$ is the Weierstrass function, and the sum is over all
positive roots $\vec{\alpha}$.
Note that natural variables, $Y_i$, correspond to the simple roots
${\vec{\alpha}_i}$, \ie
$$Y_i = {\vec{\alpha_i}} \cdot {\vec Y}\,.$$
The Weierstrass function has an expansion
\eqn\WeierPot{{\cal P}(Y) = \sum_{k=1}^{\infty} k\,  \exp(- k Y)~+~
\sum_{k,n=1}^{\infty} \, k\,  q^{kn}[\exp(-kY) +\exp(kY)- 2]\,,}
where $q = exp(-\tau)$, and where we have subtracted a constant
and rescaled to remove some factors of $4 \pi^2$.   Since the chiral
field $Y_j$ is the
action of the three-dimensional instanton in the corresponding $U(1)$,
the superpotential \ds\ is a sum over instantons.
More precisely, $Y_j = \sigma_{j} + \tau \theta_{j}$ where
$\theta_{j}$ is the holonomy of the flat connection
on ${\bf S^1}$ which breaks the gauge group to $U(1)^r$, and
${\sigma_j}$
are duals of the three-dimensional photons. The three-dimensional
instantons come from the monopoles of the four-dimensional theory whose
action is
given by $Y_j$'s, and four-dimensional instantons whose action is
$\tau$.

Our purpose here is to show that the results for the superpotential,
obtained by perturbative means in the previous sections,
are compatible with the results of \refs{\DoreySJ ,\dohol}. We expect
the agreement of terms up to $S^{h}$ order, as it is these terms that
are unambiguously computed by perturbation theory.
In fact, for the gauge group $U(N)$, this was already checked in
\refs{\dv,\Doreya} ,
so here we will generalize this to arbitrary gauge groups.

First, note that in the limit of large mass, we should recover
the affine-Toda superpotential.   This is because in that limit the
theory becomes
pure ${\cal N}=1$ SYM and moreover, the variables $Y_j$ are
{\it precisely} the same  -- the vector-scalar duality of the
three-dimensional
theory on a circle is the two-dimensional  mirror symmetry which we
used to
relate the moduli space of flat connections on $T^2$ to the affine-Toda
theory.
%
Recalling the results of the previous section is
natural to define $Y_0$ through the affine-Toda constraint
\eqn\todaconstr{\tau~=~  \sum_{a=0}^{r} \, p_a \, Y_a\,,}
where $\sum_{i=1}^r {p_i} {\alpha_i}$ is the highest root of the Lie
algebra, and $p_0 =1$.
%
%
%
%

To show that the superpotentials coincide, we proceed as follows.
We write the superpotential $W(\vec{Y};\tau)$
as a superpotential $\cW(Y) $
that depends upon the $Y_a =\{Y_i;~Y_0\}$ by using the affine-Toda
constraint
to eliminate $\tau$. This does not change the theory
provided we introduce a Lagrange multiplier field $S$ which imposes the
affine-Toda constraint:
\eqn\Wnew{W(Y;S)  = \cW(Y)  ~-~   S \Big(\tau ~-~
   \sum_{a=0}^{r} \, p_a Y_a\Big)\,.}
The Lagrange multiplier field $S$ must be
identified with the glueball superfield, for the same reason
as in the affine-Toda case. To get the glueball superpotential
one simply integrates out  the $Y$'s.

This can easily be done perturbatively,
as there is a natural $q$-grading of roots.
This is because, on the one hand, there is a grading corresponding
to the charge lattice of instantons, generated by the
instantons of smallest actions corresponding to $Y_a$'s and $Y_0$
themselves,
and on the other hand, in the confining vacuum, to the leading order, these
scale as $e^{-Y_a} \sim q^{1/h}$.
Note that in keeping only the contributions to \ds\ coming
{}from the monopoles of charges corresponding to the simple roots of the
affine Lie algebra, one precisely recovers the affine-Toda
superpotential to leading order.

The superpotential \ds\ has an expansion of the form:
\eqn\Wexpansion{\cW(Y) = \sum_a y_a ~+~  \sum_{\ell=2}^\infty
\ \
\sum_{a_1,\dots, a_\ell=0}^r  \,B^{(\ell)}_{a_1,\dots,a_\ell}  \,
y_{a_1} \dots y_{a_\ell} \,,}
where we define $y_a = e^{-Y_a}$ (For simplicity, we have set the mass,
$m$,  to one). A term of degree $\ell$ in the $y_a$ corresponds to a
 term that scales
$q^{\ell/h}$, and is thus related to the  number, ${\ell}$, of loops in the
perturbative calculation.
The one-loop term was computed in the previous section.  At two loops,
we should include monopoles whose  charge corresponds to two times a simple
(affine) root, and thus corresponds to a (non-affine) root  that can
 be written as the
sum of two simple affine roots.  For simply-laced Lie algebras, this means
that the affine simple roots must have an inner product of $-1$.  This is
precisely what is encoded in the extended Dynkin diagram of the Lie algebra
, and
indeed, one has:
\eqn\matdefs{B^{(2)}_{a b} ~=~  3\, \delta_{ab} - \coeff{1}{2}
\,{\widehat C}_{ab} ~
\equiv~ 2\, \delta_{ab}  +  \coeff{1}{2} \, \cI_{ab} \,,}
where ${\widehat C}$ is the extended Cartan matrix, and $\cI$ is
the incidence matrix of the extended diagram.   So long as one has
$\ell <h$ the expansion \Wexpansion\ is entirely characterized by this
incidence matrix, and in terms of walks of a certain length on the
extended
Dynkin diagram.  For example,
$$B^{(3)}_{a b c} ~=~  \coeff{1}{2} \, \big(\cI_{(ab} \, \cI_{bc)} ~-~
   \delta_{(ab}  \, \cI_{bc)} \big) \,, \qquad {\rm with \ no \ sum \ on}
\  b \,.$$

To see how the result \Srepl, \Srepldef\ emerges here it is convenient
to introduce more ``auxilliary glueball fields.''  That is, we
introduce a dual variable, $S_a$, for each of the $Y_a$'s, making
sure, at each step, that the superpotentials contain equivalent
data. Introduce $2 (r+1)$ new variables $S_a$ and ${\hat Y}_a$, and
write the superpotential as
$$W= \cW( {\hat Y}) -  S\Big(\tau -
\sum_{b=0}^r  \, p_a \,Y_a \Big) ~+~ \sum_{a=0}^r  S_a \, ({\hat
Y}_a-Y_a)  \,.$$
Integrating out the $S_a$ sets ${\hat Y}_a= Y_a$, and the result
reduces to \Wnew.
However, one could instead integrate out the ${\hat Y}_a$  which gives:
\eqn\finalW{W=W_D(S_a) ~-~  \sum_{b}\, Y_a\,(S_b ~-~ p_b \,S) ~-~ \tau
S\,.}
The $W_D(S_a)$ is the Legendre transform of  $\cW(Y_a)$:
$$ W_D(S_a) = \{\cW( {\hat Y}_a) - \sum_b S_b {\hat Y}_b\}_{S_a =
- {\del \cW \over \del {\hat Y}_a}}, \, $$
and as such it contains exactly the same information  as
the original superpotential in \ds.   In particular, the
Legendre transformation can
be inverted to recover $\cW(Y_a)$ from $W_D(S_a)$.
Finally, note that the superpotential, $W$ in \finalW\ has Lagrange
multipliers $Y_a$ that impose the condition:
\eqn\Selim{ S_a  ~=~  p_a \, S\,.}
%
%
One can easily show that to third order one has:
\eqn\WSexp{ \eqalign{ W  ~=~    \sum_{a=0}^r  \, S_a (\log(S_a) -1) &~-~
\sum_{a, b =0}^r  \, \big( 2\, \delta_{ab}  +  \coeff{1}{2} \, \cI_{ab}
\big)
    \, S_{a}\, S_{b}\cr &  ~+~  \coeff{1}{2} \,
\sum_{a, b,c =0}^r  \, \big(  10\,  \delta_{ab} \, \delta_{bc} +  9\,
   \delta_{ab}  \, \cI_{bc} \big) \, S_{a}\, S_{b} \, S_{c}
~+~\cO(S^4)\,,}}
where we have made use of the fact that the entries of the incidence
matrix satisfy: $ \cI_{ab} \cI_{ba}= \cI_{ab}$ (with no sum on the
indices). Using \Selim\ and recalling that the $p_a$ are
a null vector of $\widehat C_{ab}$, one finds:
\eqn\WSps{ \eqalign{W  ~=~  & h\,  S \, (\log(S) -1)  -   3 \,
\Big( \sum_{a=0}^r \, (p_a)^2 \Big) \, S^2  \cr   & +   14\,
\Big( \sum_{a =0}^r \, (p_a)^3 \Big) \, S^3  -115\,
\Big( \sum_{a =0}^r \, (p_a)^4 \Big) \, S^4 ~ +~\cO(S^5)\,,}}
where we have dropped a term $S\, \sum_a p_a \log p_a$ which can
be absorbed to the definition of $\tau$.   Here
the $S^4$ term has been obtained from the result for $U(N)$
and the observation that the fourth order result can be written entirely
in terms of $\cI_{ab}$ and $\delta_{ab}$,  and so must be proportional
to  $\sum_{a =0}^r \, (p_a)^4$.
It is also easy to convince oneself that this structure
continues, namely the terms in $W_D(S)$ can be
expressed solely in terms of $\cI_{ab}$ and $\delta_{ab}$. For this,
it is useful to
use the eigenvectors of $\widehat C$, and the result is an expansion
in powers of the entries of such eigenvectors.
What is important is that \Selim\ is precisely
the null vector of $\widehat C$, and hence
is an eigenvector of $\cI$ with eigenvalue $2$.
This gives further support for our
conjecture
at the end of section 6.

Note that it may be tempting to try to relate the $S_a$'s as
glueball superfields of
the abelian background (up to rescaling by $p_a$'s):  $S_a = W_a^2$.
Even though this should morally be correct, there are some subtleties
to understand:
the naive identification would lead one to expect the
$\ell$-loop
contribution  of the form $(\sum_a S_a)^{\ell}$, which is not the case.

\newsec{Final Comments}

In this paper we have seen that the computation of glueball
superpotential can be carried out for all groups
and representation for low powers of the glueball field explicitly
and unambiguously.  For higher powers we have discussed the existence
of an ambiguity which relates to a UV complete definition of F-terms.
For classical groups (and in some
cases for non-classical groups) we have found a way to resolve the ambiguity
by embedding the theory as a higgs branch of an arbitrarily
large rank supergroup.  Moreover we have discussed how
this agrees or differs from more standard UV completions
for some examples.  It would be very interesting
to find ``all possible physically consistent F-term completions''
and how they related to one another. This reminds one of the
framing ambiguity of Chern-Simons theory, which is needed
to complete the quantum definition of the theory \WittenHF.
It also reminds one of coupling topological matter to
topological gravity in $(1+1)$ dimensions: That is, one
adds new physical observables, topological gravity, to extend
and complete the topological free energy.  This extension
can then be computed from the ``small phase space'' of
the topological matter via differential recurrence relations.
It is thus tempting to conjecture that a given  F-completion
is equivalent to the choice of a ``reduced phase space''
inside this infinite dimensional phase space.  It would be
very interesting
to characterize all such physically consistent choices and
what relations this leads to for the computation
of the coefficients of the glueball superpotential.

Another direction, currently under investigation, is to find alternative
matrix models, which account for the possible residual instanton 
contributions to superpotentials.  The occasional discrepancies associated
with residual instantons effects can themselves be captured by a matrix
model, within the original matrix models.  One can perhaps regard this
as ``matrix model epicycles''.  

\bigskip
\leftline{\bf
   Acknowledgements}

We would like to thank A. Brandhuber, R. Dijkgraaf, J. Gomis,
M. Grisaru, P. Kraus, H. Ooguri, C. Romelsberger, J. Schwarz and M. Shigemori 
for valuable discussions.

 The research of MA and CV was supported in part by NSF grants
PHY-9802709 and DMS-0074329.  The work of KI was supported by DOE
grant DOE-FG03-97ER40546.  The work of NW was supported in part by
funds provided by the DOE under grant number DE-FG03-84ER-40168.
CV thanks the hospitality of the theory group at Caltech,
where he is a Gordon Moore Distinguished Scholar.
MA is also gratefull to theory group at Caltech for hospitality
during a part of this work.

\appendix{A}{A Weyl-invariant tensor}

Invariants defined on the whole Lie Algebra are equivalent  to
Weyl-invariant tensors defined on the Cartan subalgebra.  The standard
Casimir
invariants reduce to {\it symmetric}, Weyl-invariant polynomials on the
Cartan
subalgebra, and the classification of such invariants is well-known.

In the computation of the superpotential we encountered a slightly
different
tensor on the Cartan subalgebra:
$$
T_{\mu_1 \mu_2 \dots \mu_\ell}^ {\nu_1 \nu_2 \dots \nu_\ell} \,,
$$
satisfying the conditions:
\item{(i)}  $T$ is completely skew in ${\mu_1 \mu_2 \dots \mu_\ell}$,
\item{(ii)}  $T$ is completely  skew in ${\nu_1 \nu_2 \dots \nu_\ell}$,
\item{(iii)}  $T$ is Weyl invariant.

Our purpose here is to show that such a tensor must have the form:
\eqn\SUNform{
T_{\mu_1 \mu_2 \dots \mu_\ell}^ {\nu_1 \nu_2 \dots \nu_\ell} ~=~ const.
~
\delta^{[\nu_1}_{[\mu_1} \, \delta^{\nu_2}_{\mu_2} \, \dots\,
\delta^{\nu_\ell]}_{\mu_\ell]} \,,}
and hence, when contracted with $W$'s gives only an $S^\ell$ term.

We start by considering $SU(N)$, whose Weyl group is $S_N$, the
permutation
group on $N$ elements.   It is more convenient to go to $U(N)$ and
introduce
the standard, orthonormal weight basis, $e_1,\dots, e_N$.  The cost of
doing this
is that there is a natural, Weyl-invariant vector, $V\equiv  e_1 + e_2 +
\dots + e_N$,
which defines the overall $U(1)$ factor in $U(N)$.

In this instance we will show that:
\eqn\UNform{
T_{\mu_1 \mu_2 \dots \mu_\ell}^ {\nu_1 \nu_2 \dots \nu_\ell} ~=~
a \, \delta^{[\nu_1}_{[\mu_1} \, \delta^{\nu_2}_{\mu_2} \, \dots\,
\delta^{\nu_\ell]}_{\mu_\ell]} ~+~
b\,  \delta^{[\nu_1}_{[\mu_1} \, \delta^{\nu_2}_{\mu_2} \, \dots\,
\delta^{\nu_{\ell-1}}_{\mu_{\ell-1}} \,
V^{\nu_\ell]}\, V_{\mu_\ell]}  \,,}
for some constants $a$ and $b$.  This establishes the $SU(N)$ result
once
\UNform\ is projected onto the $SU(N)$ subalgebra.

The form of the argument is most easily seen by starting with $\ell =1$,
with a
tensor  $T_\mu^\nu$.  Suppose that $ T^1_2 \ne 0$ and define:
$$
\widetilde T_\mu^\nu \equiv T_\mu^\nu ~-~ T^1_2  \,   V^\nu \,
V_\mu \,.
$$
By construction, $\widetilde T^1_2 =  0$.  Acting with the Weyl group
shows
that all the $\widetilde T_\mu^\nu$ must be zero for $\nu \ne \mu$.
Consider $\widetilde T_\mu^\mu$ (no sum on indices): acting with the
Weyl group implies that all of these elements must be equal. Thus
$\widetilde T_\mu^\nu =  \delta_\mu^\nu$, and the result follows.

If $\ell >1$, then the indices  $\{\mu_1,\dots, \mu_\ell\}$  must be
distinct,
and so must the indices  $\{\nu_1,\dots, \nu_\ell\}$.  Moreover,  {\it
at most} one of the
$\mu_k$ does not lie in the set $\{\nu_1,\dots, \nu_\ell\}$.  Otherwise,
if $\mu_p$ and
$\mu_q$ are distinct from all the $\{\nu_1,\dots, \nu_\ell\}$, apply the
permutation
that interchanges $\mu_p$ and $\mu_q$:  Skew symmetry in the
$\mu$'s and Weyl invariance then imply
$$
T_{\mu_1 \mu_2 \dots \mu_\ell}^ {\nu_1 \nu_2 \dots \nu_\ell} ~=~ -
T_{\mu_1 \mu_2 \dots \mu_\ell}^ {\nu_1 \nu_2 \dots \nu_\ell}  \,,
$$
and hence this component is zero.

Let $\rho_1\,, \dots, \rho_\ell$ and $\sigma_1\,, \dots, \sigma_\ell$ be
one set of values
of the indices  for which  $T_{\rho_1 \rho_2 \dots \rho_\ell}^ {\sigma_1
\sigma_2 \dots
\sigma_\ell} \ne 0$ and yet  $\{\rho_1\,,\dots, \rho_\ell\}$ is not a
just a permutation of
$\{\sigma_1\,, \dots, \sigma_\ell\}$.  Define:
$$
\widetilde T_{\mu_1 \mu_2 \dots \mu_\ell}^ {\nu_1 \nu_2 \dots
\nu_\ell} ~=~ T_{\mu_1
\mu_2 \dots \mu_\ell}^ {\nu_1 \nu_2 \dots \nu_\ell}  - b\,
\delta^{[\nu_1}_{[\mu_1} \, \delta^{\nu_2}_{\mu_2} \, \dots\,
\delta^{\nu_{\ell-1}}_{\mu_{\ell-1}} \,
V^{\nu_\ell]}\, V_{\mu_\ell]} \,,
$$
and since only one of the $\rho$'s does not lie in the
set of $\sigma$'s, it is possible to choose $b$ so that
$\widetilde T_{\rho_1 \rho_2 \dots \rho_\ell}^ {\sigma_1 \sigma_2 \dots
\sigma_\ell} = 0$.  By acting with the permutation group one sees
that the only non-zero elements of
$\widetilde T_{\mu_1 \mu_2 \dots \mu_\ell}^ {\nu_1 \nu_2 \dots
\nu_\ell} $ are
those in which $\{\mu_1,\dots, \mu_\ell\}$  are a permutation
of the indices  $\{\nu_1,\dots, \nu_\ell\}$, and moreover all these
tensor components must be equal.  The result follows.

Once one has the result for $SU(N)$ it is relatively trivial for the
other groups:
One simply goes to a big enough $U(M)$ subgroup.  For $G_2$, $E_7$ and
$E_8$ it is trivial since they have $SU(3)$, $SU(8)$ and $SU(9)$ as
maximal
subgroups.  The result then follows from the result for $SU(N)$.
Every other Lie algebra of rank $r$ has a maximal subalgebra of
$U(r)$.  From this we can deduce that $T$ must have the form
\UNform, however for a semi-simple Lie algebra there are no
fixed vectors, $V$, and so the coefficient, $b$, must be zero.

\vfill
\eject

\listrefs

\end